%% file: main.tex
\documentclass[aps,prd,reprint,preprintnumbers,superscriptaddress,amsmath,amssymb,floatfix,longbibliography]{revtex4-1}

\usepackage{graphicx}
\usepackage{epsfig}
\usepackage{dcolumn}
\usepackage{bm}
\usepackage{multirow}
\usepackage[colorlinks,linkcolor=red,anchorcolor=green,citecolor=blue,CJKbookmarks=True]{hyperref}

\usepackage{tikz, tikz-feynhand}
\usetikzlibrary{shapes.geometric, arrows}

\begin{document}
\title{$X(3872)$ relevant $D\bar{D}^*$ scattering in $N_f=2$ lattice QCD}

\author{Haozheng Li}
\email{haozhengli2002@gmail.com}
\affiliation{Institute of High Energy Physics, Chinese Academy of Sciences, Beijing 100049, People's Republic of China}
\affiliation{School of Physics and Technology, Wuhan University, Wuhan 430072, People's Republic of China}

\author{\small Chunjiang Shi}
\email{shichunjiang@ihep.ac.cn}
\affiliation{Institute of High Energy Physics, Chinese Academy of Sciences, Beijing 100049, People's Republic of China}
\affiliation{School of Physical Sciences, University of Chinese Academy of Sciences, Beijing 100049, People's Republic of China}

\author{Ying Chen}
\email{cheny@ihep.ac.cn}
\affiliation{Institute of High Energy Physics, Chinese Academy of Sciences, Beijing 100049, People's Republic of China}
\affiliation{School of Physical Sciences, University of Chinese Academy of Sciences, Beijing 100049, People's Republic of China}

\author{Ming Gong}
\affiliation{Institute of High Energy Physics, Chinese Academy of Sciences, Beijing 100049, People's Republic of China}
\affiliation{School of Physical Sciences, University of Chinese Academy of Sciences, Beijing 100049, People's Republic of China}

\author{Juzheng Liang}
\affiliation{Institute of High Energy Physics, Chinese Academy of Sciences, Beijing 100049, People's Republic of China}
\affiliation{School of Physics, University of Science and Technology of China, Hefei 230026, People’s Republic of
    China}

\author{Zhaofeng Liu}
\affiliation{Institute of High Energy Physics, Chinese Academy of Sciences, Beijing 100049, People's Republic of China}
\affiliation{Center for High Energy Physics, Peking University, Beijing 100871, People's Republic of China}

\author{Wei Sun}
\affiliation{Institute of High Energy Physics, Chinese Academy of Sciences, Beijing 100049, People's Republic of China}

\begin{abstract}
    We study the $S$-wave $D\bar{D}^*(I=0)$ scattering at four different pion masses $m_\pi$ ranging from 250 MeV to 417 MeV from $N_f=2$ lattice QCD. Three energy levels $E_{2,3,4}$ are extracted at each $m_\pi$. The analysis of $E_{2,3}$ using the effective range expansion (ERE) comes out with a shallow bound state below the $D\bar{D}^*$ threshold, and the phase shifts at $E_{3,4}$ indicate the possible existence of a resonance near 4.0 GeV. We also perform a joint analysis to $E_{2,3,4}$ through the $K$-matrix parameterization of the scattering amplitude. In this way, we observe a $D\bar{D}^*$ bound state whose properties are almost the same as that from the ERE analysis. At each $m_\pi$, this joint analysis also results in a resonance pole with a mass slightly above 4.0 GeV and a width around 40-60 MeV, which are compatible with the properties of the newly observed $\chi_{c1}(4010)$ by LHCb. More scrutinized lattice QCD calculations are desired to check the existence of this resonance.

\end{abstract}
\maketitle
\section{Introduction}
Ever since its discovery in 2003~\cite{Belle:2003nnu}, $X(3872)$ (aka $\chi_{c1}(3872)$~\cite{ParticleDataGroup:2022pth}) has been a hot topic in experimental and theoretical studies. It has quantum numbers $I^G J^{PC}=0^+ 1^{++}$~\cite{LHCb:2013kgk}, a mass almost at the $D^0\bar{D}^{0*}$ threshold, and a very tiny width of $\sim 1~\mathrm{MeV}$. It decays mainly into $D^0\bar{D}^{0*}$, and has also decay modes $J/\psi \omega$ (I=0) and $J/\psi \rho^0$ (I=1) with comparable branching fractions~\cite{ParticleDataGroup:2022pth}. The exotic properties of $X(3872)$ imply that it is not a pure $\chi_{c1}(2P)$ charmonium whose mass is expected to be above 3.9 GeV~\cite{Godfrey:1985xj,Barnes:2005pb}. Phenomenological studies also interpret it to be a loosely bound $D\bar{D}^*$ molecule~\cite{Tornqvist:2004qy,Liu:2008fh,Liu:2009qhy,Li:2012cs,Qiu:2023uno,Dong:2021juy} (and possibly a small fraction of the $c\bar{c}$ component~\cite{Braaten:2003he,Kalashnikova:2005ui,Barnes:2007xu,Ortega:2009hj,Li:2009ad,Baru:2010ww,Yamaguchi:2019vea}), or a compact tetraquark state~\cite{Maiani:2004vq,Ebert:2005nc}. See Refs.~\cite{Brambilla:2010cs,Chen:2016qju,Esposito:2016noz,Guo:2017jvc,Brambilla:2019esw,Liu:2019zoy} for reviews.

$X(3872)$ relevant lattice QCD studies observe a strong coupling between $c\bar{c}$ and $D\bar{D}^{(*)}$~\cite{Bali:2011rd, Prelovsek:2013cra, Padmanath:2015era}. This coupling can result in a bound state corresponding to $X(3872)$ below the $D\bar{D}^*$ threshold~\cite{Prelovsek:2013cra,Padmanath:2015era}. However, the study in Ref.~\cite{Prelovsek:2013cra} does not pay enough attention to the role played by the possible $\chi_{c1}(2P)$ in interpreting the energy levels above the $D\bar{D}^*$ threshold. On the other hand, the $D\bar{D}^*$ interaction can be mediated by light hadrons and therefore can be sensitive to light quark masses. So we will revisit the $X(3872)$ relevant $D\bar{D}^*$ scattering at different pion masses and explore the quark mass dependence of the $D\bar{D}^*$ interaction, which may provide more information to the existing effective field theories describing the $D\bar{D}^*$ interaction~\cite{AlFiky:2005jd,Albaladejo:2013aka,Baru:2013rta,Wang:2013kva,Guo:2017jvc}. In lattice QCD, the $D\bar{D}^*$ scattering amplitude can be derived from the related finite volume energy levels through L\"{u}scher method~\cite{Luscher:1986pf,Luscher:1990ux,Luscher:1991cf}. For the $X(3872)$ relevant $D\bar{D}^*$ scattering, the major numerical challenge is the calculation of the light quark annihilation diagrams that contribute to the related correlation functions. For this purpose, we adopt the distillation method~\cite{Peardon:2009gh} that provides a sophisticated treatment of both the all-to-all quark propagators and a smearing scheme for quark fields.
\begin{table*}[t]
    \caption{Parameters of the gauge ensembles and the energies of $D$ and $D^*$ with $q=0,1$. The non-interacting $D\bar{D}^*$ energies $E_{D\bar{D}^*}^{q=0,1}$ are also shown.}
    \label{tab:config-m}
    \begin{ruledtabular}
        \begin{tabular}{lcccccc|ccccccc}
            ens.            & $m_\pi$         & $a_t^{-1}$             & $N_\mathrm{cfg}$ & $N_{V}^{(\mathrm{l})}$ & $N_V^{(\mathrm{c})}$ &
            $m_{\chi_{c1}}$ &
            $m_D$           & $m_{D^*}$       & $E_{D\bar{D}^*}^{q=0}$ &
            $E_D^{q=1}$     & $E_{D^*}^{q=1}$ & $E_{D\bar{D}^*}^{q=1}$                                                                                                                                          \\
                            & (MeV)           &                        &                  &                        &                      & (MeV)   & (MeV)   & (MeV)   & (MeV)   & (MeV)   & (MeV)   & (MeV)   \\\hline
            M245            & 250(3)          & $7.276$                & 401              & 70                     & 120                  & 3489(3) & 1873(1) & 1985(2) & 3858(3) & 1958(2) & 2064(4) & 4022(5) \\
            M305            & 307(2)          & $7.187$                & 401              & 70                     & 120                  & 3496(2) & 1881(1) & 1990(2) & 3871(3) & 1962(2) & 2070(2) & 4032(4) \\
            M360            & 362(1)          & $7.187$                & 401              & 70                     & 120                  & 3502(2) & 1884(1) & 2003(2) & 3888(2) & 1970(1) & 2084(3) & 4054(4) \\
            M415            & 417(1)          & $7.219$                & 401              & 70                     & 120                  & 3509(2) & 1896(1) & 2017(1) & 3913(2) & 1978(1) & 2094(2) & 4072(3) \\
        \end{tabular}

    \end{ruledtabular}
\end{table*}

\section{Numerical details}
We generate $N_f=2$ gauge configurations at four pion masses ranging from 250 MeV to 417 MeV on $16^3\times 128$ anisotropic lattices. The aspect ratio is tuned to be $\xi=a_s/a_t\approx 5$, where $a_s$ and $a_t$ are the spatial and temporal lattice spacings, respectively.
We adopt the tadpole improved gauge action~\cite{Morningstar:1997ff,Chen:2005mg} for gluons and the tadpole improved anisotropic clover fermion action~\cite{Zhang:2001in,Su:2004sc,CLQCD:2009nvn} for light sea quarks and the valence charm quark.
The spatial lattice spacing $a_s$ is set to be $a_s=0.136(2)~\mathrm{fm}$ through the Wilson flow method~\cite{Luscher:2010iy,BMW:2012hcm}.
Then the pion dispersion relation along with the measured $\xi$ are used to determine $a_t$ at each pion mass.
The charm quark mass parameter is tuned to give $(m_{\eta_c}+3m_{J/\psi})/4=3069$ MeV. Throughout this study, the correlation functions are calculated using the distillation method~\cite{Peardon:2009gh}. We use $N_V^{(\mathrm{l})}=70$ in calculating the perambulators for $u,d$ quarks, and $N_V^{(\mathrm{c})}=120$ in the calculation of charm quark perambulators, where $N_V$ is the number of the eigenvectors of the gauge covariant Laplacian operators on the lattice. The parameters for the gauge ensembles are listed in Table~\ref{tab:config-m}.

We only consider the possible coupled-channel effects of $\chi_{c1}$ states and the $S$-wave $D\bar{D}^*$ and ignore the coupling from the $J/\psi\omega$ system, which is observed to be very weak (see Ref.~\cite{Prelovsek:2013cra} and also Appendix~\ref{sec:appendix-S1} and \ref{sec:appendix-S2}). {Note that the $D$-wave ($l=2$) should also appear
        in the related $D\bar{D}^*$ scattering. However, its mixing to the $S$-wave is expected to be suppressed by the phase space factor $p^{2l}$ with $p$ being the scattering momentum and is assumed to be negligible in the practical study~\cite{Prelovsek:2013cra}.} The related interpolating field operators are built in terms of the smeared $u,d,c$ quark fields. The $c\bar{c}$ operators take the spatially extended version,
$\mathcal{O}_{c\bar{c}}^{r}(t)=\frac{1}{N_r}\sum\limits_{|\mathbf{y-x}|=r}\bar{c}(\mathbf{x},t) \gamma_5\gamma_i K_U (\mathbf{x},\mathbf{y};t)c(\mathbf{y},t)$,
where $N_r$ is the multiplicity of $\mathbf{r}=\mathbf{y-x}$ with $|\mathbf{r}|=r$, and
$   K_U(\mathbf{x},\mathbf{y};t)=\mathcal{P}e^{ig\int_\mathbf{y}^\mathbf{x}\mathbf{A}\cdot d\mathbf{r}}$
is the Wilson line connecting $(\mathbf{y},t)$ and $(\mathbf{x},t)$ (The spatial index $i$ of all the vector operators is omitted for convenience throughout this Letter).
Obviously, $\mathcal{O}_{c\bar{c}}^r(t)$ is gauge invariant and has the right quantum number $J^{PC}=1^{++}$ after the summation over $|\mathbf{r}|=r$. In practice, we use three $\mathcal{O}_{c\bar{c}}^r(t)$ operators (denoted as $\mathcal{O}_{1,2,3}$ in the following) with $r/a_s=0,1,2$, respectively.

The $D\bar{D}^*$ operators with the quantum numbers $I^G J^{PC}=0^+ 1^{++}$ take the meson-meson operator type
\begin{equation}
    \mathcal{O}_{D\bar{D}^{*}} \sim \left[(\bar{u}\Gamma_D c)(\bar{c}\Gamma_{D^*}u) - (\bar{u}\Gamma_{D^*}c)(\bar{c}\Gamma_D u)\right]+(u\to d),
\end{equation}
with $(\Gamma_D,\Gamma_{D^*})=(\gamma_5, \gamma_i)$. $S$-wave operators are obtained by summing over all the possible directions of the relative momentum, namely,
$\mathcal{O}_{AB}^q\sim \sum\limits_{R\in O} \mathcal{O}_A(R\circ\mathbf{q})\mathcal{O}_B(-R\circ \mathbf{q})$,
where $R$ runs over all the elements of the octahedral group $O$, and $q=|\mathbf{q}|$ is the magnitude of the relative momentum (in units of $2\pi/L$ with $L$ being the spatial size of the lattice).
The operators $\mathcal{O}_{4, 6}\equiv \mathcal{O}_{D\bar{D}^*}^{q=0, 1}$ are involved in the practical calculation. Another $\mathcal{O}_{D\bar{D}^*}^{q=0}$ operator (labeled as $\mathcal{O}_5$) uses the combination $(\Gamma_D,\Gamma_{D^*})=(\gamma_4\gamma_5, \gamma_4\gamma_i)$.

\subsection{Finite volume energies}
The six operators $\{\mathcal{O}_\alpha, \alpha=1,2,\ldots,6\}$ are used to calculate the correlation matrix
$C_{\alpha\beta}(t)=\langle 0|\mathcal{O}_\alpha(t)\mathcal{O}^\dagger_\beta(0)|0\rangle$.
The practical calculation considers all the quark diagrams, except for those involving the charm quark annihilation. By solving the generalized eigenvalue problem (GEVP),
$C_{\alpha\beta}(t)v_\beta^{(n)}(t,t_0)=\lambda^{(n)}(t,t_0) C_{\alpha\beta}(t_0) v_\beta^{(n)}(t,t_0)$
at given $t$ and $t_0$, we obtain the optimized correlation function $C^{(n)}(t)=v^{(n)}_\alpha v^{(n)}_\beta C_{\alpha\beta}(t)$ which is contributed mainly by the $n$-th state. The finite volume energies $E_{2,3,4}$ are extracted through one-exponential fits to the ratio functions~\cite{CP-PACS:2007wro,BaryonScatteringBaSc:2023ori,BaryonScatteringBaSc:2023zvt}
\begin{equation}\label{eq:ratio}
    R_n(t)=\frac{C^{(n)}(t)}{C_D(t,q)C_{D^*}(t,q)}\approx A e^{-\Delta_n t},
\end{equation}
where $C_{D^{(*)}}(t,q)$ is the correlation function of $D^{(*)}$ with a momentum $q=0, 1$, and $\Delta_n=E_n-E_{D\bar{D}^*}^q$ is the difference of $E_n$ from the nearest non-interacting $D\bar{D}^*$ energy. In practice, we use $q=0$ for $E_{2,3}$ and $q=1$ for $E_4$. We also perform two-exponential fits to $C^{(n)}(t)$ in proper time windows for a self-consistent check in Appendix~\ref{sec:appendix-S3}. All the statistical errors are obtained through the jackknife method.

\begin{figure}[t]
    \centering
    \vspace{-0.15in}
    \includegraphics[width=0.98\linewidth]{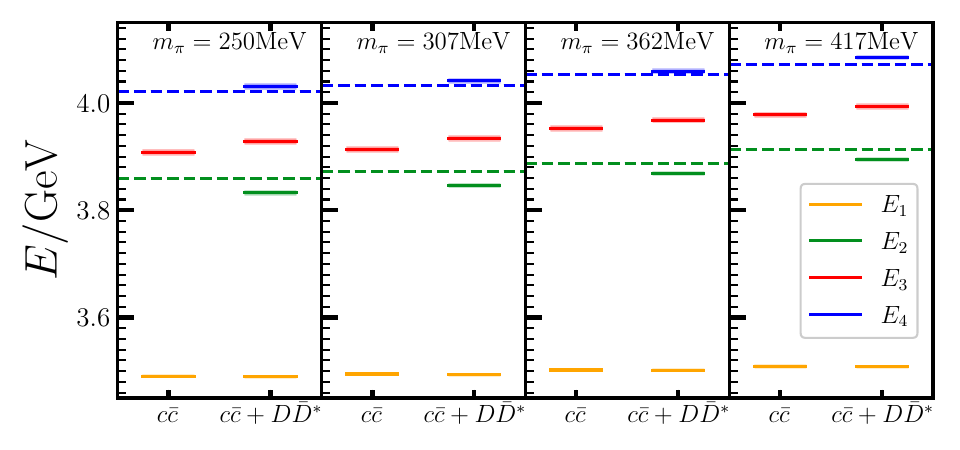}
    \vspace{-0.20in}
    \caption{Energy levels $E_n$ obtained at different $m_\pi$. In each panel, the dashed lines show the non-interacting $D\bar{D}^*$ energies $E_{D\bar{D}^*}^{q=0}$ (green) and $E_{D\bar{D}^*}^{q=1}$ (blue) in Table~\ref{tab:config-m}, while the boxes stand for the finite volume energy levels $E_n$ with their heights indicating statistical errors. In the left column of each panel are the lowest two $E_n$ when only $c\bar{c}$ operators are involved in the GEVP analysis, and the right column includes the lowest four energy levels $E_n$ from all the six operators ($\mathcal{O}_{1-6}$).
        It should be noted that there may exist an additional energy level corresponding to the $D$-wave ($l=2$) $D\bar{D}^*$ state near the blue dashed line in the figure. Similar to the treatment in Ref.~\cite{Prelovsek:2013cra}, we tentatively neglect the mixing from the $D$-wave wave state to the $S$-wave, which is expected to be suppressed by the phase space factor $p^{2l}$ particularly for the low-energy $D\bar{D}^*$ scattering~\cite{Padmanath:2022cvl}. The effect of this kind of mixing needs to be explored by involving more $D\bar{D}^*$ operators.}
    \vspace{-0.20in}
    \label{fig:diagram_gevp}
\end{figure}

The obtained finite volume energy levels are illustrated in Fig.~\ref{fig:diagram_gevp} by colored boxes, whose heights indicate the statistical errors. In each panel labeled by $m_\pi$, the yellow and red boxes in the left column are the two energy levels that are derived from the GEVP analysis involving only the $c\bar{c}$ operators ($\mathcal{O}_{1,2,3}$). The lowest level is roughly 3.5 GeV and corresponds to the conventional charmonium $\chi_{c1}$, while the second level is higher than 3.9 GeV and coincides with the quark model expectation of the $\chi_{c1}(2P)$ mass~\cite{Godfrey:1985xj,Barnes:2005pb,Ebert:2002pp,Li:2009zu,Deng:2016stx,Wang:2019mhs}. Note that this energy may not be an eigenvalue of the lattice Hamiltonian. When the $\mathcal{O}_{D\bar{D}^*}$ operators $\mathcal{O}_{4,5,6}$ are added, more energy levels are obtained. The colored boxes in the right-hand side of each column in Fig.~\ref{fig:diagram_gevp} show the lowest four energy levels (labeled $E_1, E_2, E_3, E_4$ from bottom up). $E_1$ is almost the same as the lowest level obtained by $c\bar{c}$ operators. The second lowest level $E_2$ (green box) appears close to but right below the non-interacting $D\bar{D}^*$ energy $E_{D\bar{D}^*}^{q=0}$ (green dashed line), and $E_4$ lies right above the non-interacting $D\bar{D}^*$ energy $E_{D\bar{D}^*}^{q=1}$ (blue dashed line) (see Table~\ref{tab:config-m} for the exact values). It is interesting to see that $E_3$ resides in the middle of $E_{D\bar{D}^*}^{q=0}$ and $E_{D\bar{D}^*}^{q=1}$.

\begin{figure}[t]
    \centering
    \includegraphics[width=0.9\linewidth]{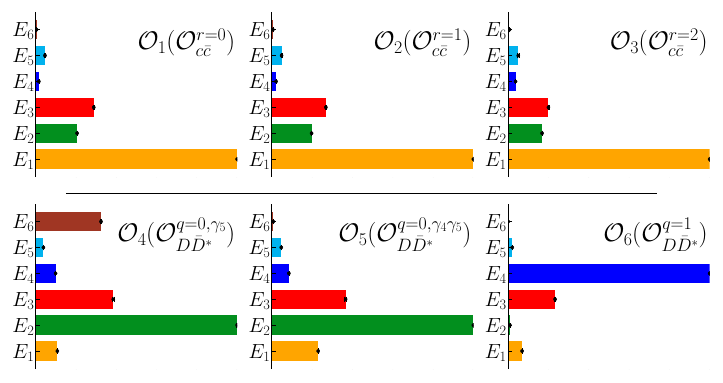}
    \caption{Relative couplings $Z_\alpha^{(n)}=\langle 0|\mathcal{O}_\alpha|n\rangle$ at $m_\pi = 417~\mathrm{MeV}$. For each operator $\mathcal{O}_\alpha$, $Z_\alpha^{(n)}$ is normalized by the largest value in $\{Z_\alpha^{(n)}, n=1,2,\ldots, 6\}$. For each state $|n\rangle$, $Z_\alpha^{(n)}$ (in the same color) signal the relative importance of $c\bar{c}$ and $D\bar{D}^*$ components.
    }
    \label{fig:overlap-weight}
    \vspace{-0.20in}
\end{figure}

$E_{1,2,3,4}$ are taken to be the eigenvalues of the lattice Hamiltonian relevant to $0^+ 1^{++}$ charmonium-like systems. The relative couplings of $\mathcal{O}_\alpha$ to different states $Z_\alpha^{(n)}=|\langle 0|\mathcal{O}_\alpha|n\rangle|$ (normalized by the largest value in each panel) are illustrated in Fig.~\ref{fig:overlap-weight}. Obviously, the $c\bar{c}$ operators $\mathcal{O}_{1,2,3}$ couple most to the $E_1$ state and also have substantial overlaps to the $E_2$ and $E_3$ states. In contrast, $\mathcal{O}_{4,5}$ ($D\bar{D}^*$ operators with $q=0$) couple mainly to $E_2$, and also overlap substantially to $E_1$ and $E_3$ states. $\mathcal{O}_6$ ($\mathcal{O}_{D\bar{D}^*}(q=1)$ operator) couples predominantly to $E_4$ and has a sizable coupling to the $E_3$ state. The $E_1$ state is naturally the $\chi_{c1}(1P)$ state. The $E_2$ and $E_3$ states are from the mixing of $\chi_{c1}(2P)$ and non-interacting $D\bar{D}^*$ state of $q=0$, which pulls $E_2$ downwards below the $D\bar{D}^*$ threshold (apart from the possible attractive interaction of $D\bar{D}^*$ due to meson exchanges) and pushes $E_3$ upwards, as exactly shown in Fig.~\ref{fig:diagram_gevp}. It is expected that $E_4$ is slightly above the non-interacting energy $E_{D\bar{D}^*}^{q=1}$ due to the mixing from $\chi_{c1}(2P)$.
It should be noted that the $D$-wave $D\bar{D}^*$ scattering state can have a $J^{PC}=1^{++}$ quantum number and is degenerate with the $S$-wave state in the non-interacting limit. The $S$-$D$-wave mixing (and also the possible different mixing from $\chi_{c1}(2P)$) may result in two energy levels near $E_{D\bar{D}^*}^{q=1}$ (the blue dashed line in Fig.~\ref{fig:diagram_gevp}). By assuming the contribution from the $D$-wave state is negligible, as indicated in the lattice QCD calculation of the $T_{cc}^+(3875)$ relevant $DD^*$ scattering~\cite{Padmanath:2022cvl, Chen:2022vpo}, we use only one operator $\mathcal{O}_{D\bar{D}^*}(q=1)$ that resembles the $S$-wave $D\bar{D}^*$ scattering state and makes the second energy level indiscernible. The effect of this kind of mixing needs to be explored by involving more $D\bar{D}^*$ operators at $q=1$.

Strictly speaking, $E_2, E_3, E_4$ are the energy levels of the $D\bar{D}^*$ scattering states in a finite box and are related to the corresponding scattering amplitude through L\"{u}scher's formula~\cite{Luscher:1986pf,Luscher:1990ux,Luscher:1991cf} for the $S$-wave scattering
\begin{equation}\label{eq:phase}
    p\cot \delta_0(p)=\frac{2}{\sqrt{\pi} L} \mathcal{Z}_{00}(1;q^2), ~~~q^2\equiv \left(\frac{L}{2\pi}\right)^2 p^2,
\end{equation}
where $p$ is the scattering momentum defined through $ E_n(p_n)=\sqrt{m_D^2+p_n^2}+\sqrt{m_{D^*}^2+p_n^2}$ for each $E_n$. Table~\ref{tab:fitresult} collects the results of $(p^2,p\cot\delta_0(p))$ of $E_{2,3,4}$ at all the four values of $m_\pi$, where the statistical errors are obtained by jackknife analyses. However, there are subtleties in the derivation of $p\cot \delta_0$ for $E_2$. The $D\bar{D}^*$ scattering may include the interaction from one pion exchange (OPE), which introduces nonanalyticity to $p\cot \delta_0(p)$ when $p^2$ is below the left-hand cut (lhc) point $(p_\mathrm{lhc}^{1\pi})^2=((m_{D^*}-m_D)^2-m_\pi^2)/4$~\cite{Du:2023hlu,Meng:2023bmz,Raposo:2023oru}.

\begin{figure*}[t]
    \centering
    \vspace{-0.15in}
    \includegraphics[width=1\linewidth]{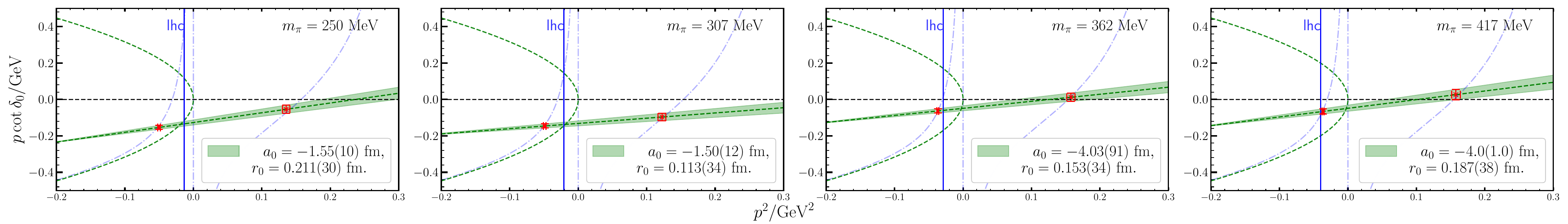}
    \caption{Phase shifts of S-wave $J^{PC}=1^{++}$ $D\bar{D}^*$ system and effective range expansion fitting.
                The red data points illustrate phase shifts $p\cot\delta_0(q)$ obtained from ($E_{2,3}$), while the green bands show the related effective ranges expansion. The background dashed lines present the Riemann-Zeta function trajectory.
        }
    \vspace{-0.20in}
    \label{fig:phase}
\end{figure*}

\begin{table*}[t]
    \caption{$D\bar{D}^*$ scattering parameters. The scattering momentum $p^2$, the phase shift $p\cot\delta_0(p)$ (and $\delta_0$) at $E_{2,3,4}$, as well as the lhc point $(p_\mathrm{lhc}^{1\pi})^2\approx \frac{1}{4}[(m_{D}-m_{D^*})^2 - m_\pi^2]$, are given explicitly at four values of $m_\pi$. The energy shifts $\Delta_n$ (derived from Eq.~(\ref{eq:ratio})) of $E_n$ from the nearby non-interacting energy of $D\bar{D}^*$ are also presented. Also collected are the derived scattering length $a_0$, the effective range $r_0$ from $E_2,E_3$, and the estimated binding energy $E_B$ ({and binding momentum $p_B=+ i|p_B|$}) for the bound state pole. The from the $K$-matrix parameters and the resulted bound state information are also shown for comparison. The asterisks in brackets `(*)' indicate that the corresponding values may be problematic when OPE lhc is considered.
    }
    \label{tab:fitresult}
    \begin{ruledtabular}
        \begin{tabular}{llrrrr}
                          & $m_\pi$(MeV)                                  & 250(3)                        & 307(2)                        & 362(1)                        & 417(1)                        \\\hline
            $E_4$         & $\Delta_4$(MeV) $=E_4 - E_{D\bar{D}^*}^{q=1}$ & 9.1(1.3)                      & 8.9(1.2)                      & 5.3(1.3)                      & 12.8(1.3)                     \\
                          & $p^2(\mathrm{GeV}^2)$                         & 0.339(8)                      & 0.335(6)                      & 0.340(6)                      & 0.342(4)                      \\
                          & $p\cot \delta_0(p)$(GeV)                      & $-$2.02(66)                   & $-$2.35(65)                   & $-$2.76(89)                   & $-$1.79(28)                   \\
                          & $\delta_0$                                    & $(163.9_{-7.3}^{+4.0})^\circ$ & $(166.1_{-5.1}^{+3.0})^\circ$ & $(168.1_{-5.5}^{+2.9})^\circ$ & $(161.9_{-3.2}^{+7.4})^\circ$ \\
            \hline
            $E_3$         & $\Delta_3$(MeV)$=E_3 - E_{D\bar{D}^*}^{q=0}$  & 70(3)                         & 63(3)                         & 80(3)                         & 80(3)                         \\
                          & $p^2(\mathrm{GeV}^2)$                         & 0.135(5)                      & 0.122(6)                      & 0.158(6)                      & 0.158(6)                      \\
                          & $p\cot \delta_0(p)$(GeV)                      & $-$0.054(19)                  & $-$0.097(19)                  & 0.012(22)                     & 0.026(24)                     \\
                          & $\delta_0$                                    & $(98.4_{-3.6}^{+3.3})^\circ$  & $(105.4_{-3.1}^{+3.0})^\circ$ & $(88.2_{-3.3}^{+3.3})^\circ$  & $(86.2_{-4.1}^{+4.0})^\circ$  \\\hline
            $E_2$         & $\Delta_2$(MeV)$=E_2 - E_{D\bar{D}^*}^{q=0}$  & $-$26.1(9)                    & $-$25.4(11)                   & $-$19.0(7)                    & $-$18.6(8)                    \\
                          & $p^2(\mathrm{GeV}^2)$                         & $-$0.050(2)                   & $-$0.049(2)                   & $-$0.037(1)                   & $-$0.036(1)                   \\
                          & $p\cot \delta_0(p)$(GeV)                      & $-$0.154(9)(*)                & $-$0.146(10)(*)               & $-$0.063(11)(*)               & $-$0.066(13)                  \\
                          & $(p_\mathrm{lhc}^{1\pi})^2 (\mathrm{GeV}^2)$  & $-$0.0135(4)                  & $-$0.0210(4)                  & $-$0.0292(3)                  & $-$0.0400(3)                  \\\hline
            ($E_{2,3}$)   & $a_{0}$ (fm)                                  & $-$1.55(10)(*)                & $-$1.50(12)(*)                & $-$4.03(91)(*)                & $\mathbf{-4.0(1.0)}$          \\
                          & $r_{0}$ (fm)                                  & 0.211(30)(*)                  & 0.113(34)(*)                  & 0.153(34)(*)                  & $\mathbf{0.187(38)}$          \\
                          & {$p_B$ ($\mathrm{MeV}$)}                      & {$+i~137(9)$(*)}              & {$+i~ 137(11)$(*)}            & {$+i~ 50(11)$ (*)}            & {$\mathbf{+i~50(13)}$}        \\
                          & $E_B$ (MeV)                                   & $-9.7^{+2.1}_{-2.2}$ (*)      & $-9.7^{+1.9}_{-2.0}$ (*)      & $-1.3^{+0.6}_{-0.8}$ (*)      & $\mathbf{-1.3^{+0.8}_{-1.0}}$ \\\hline
            {$K$-matrix}  & $M$ ($\mathrm{MeV}$)                          & 3948(6)(*)                    & 3981(11)(*)                   & 3960(11)(*)                   & $\mathbf{3979(10)}$           \\
            ($E_{2,3,4}$) & $\gamma$                                      & $-$381(22)(*)                 & $-$468(43)(*)                 & $-$750(101)(*)                & $\mathbf{-677(90) }$          \\
                          & $g(\mathrm{GeV}^2)$                           & $-$220(14)(*)                 & $-$205(25)(*)                 & $-$565(97)(*)                 & $\mathbf{-532(94)}$           \\
                          & $p_B$ ($\mathrm{MeV}$)                        & $+i~ 146(9)$(*)               & $+i~ 141(10)$(*)              & $+i~ 57(11)$ (*)              & $\mathbf{+i~58(12)}$          \\
                          & $E_B$ ($\mathrm{MeV}$)                        & $-11(1)$(*)                   & $-10(2)$(*)                   & $ -1.7(7)$(*)                 & $\mathbf{-1.7(7)}$            \\
        \end{tabular}
    \end{ruledtabular}
\end{table*}

\subsection{Bound state and the possible lhc issue}\label{sec:boundstate}
The values of $(p_\mathrm{lhc}^{1\pi})^2$ at different $m_\pi$'s are also shown in Table~\ref{tab:fitresult}. It is seen that $p^2(E_2)>(p_\mathrm{lhc}^{1\pi})^2$ is satisfied only for $m_\pi=417(1)~\text{MeV}$. So we discuss the near-threshold $D\bar{D}^*$ scattering amplitude at $m_\pi=417(1)~\text{MeV}$ following the strategy in Ref.~\cite{Prelovsek:2013cra,Mohler:2013rwa}, since the $p\cot\delta_0(p)$ derived at $E_2$ and $E_3$ from Eq.~(\ref{eq:phase}) are valid. The effective range expansion (ERE),
$p\cot \delta_0(p)= \frac{1}{a_0}+\frac{1}{2} r_0 p^2$,
at $E_2$ and $E_3$ gives two equations with two unknowns, namely, the $S$-wave scattering length $a_0$ and the effect range $r_0$. The solutions are
\begin{equation}
    a_0=-4.0(1.0)~\mathrm{fm},~~~r_0=0.187(38)~\mathrm{fm},
\end{equation}
where the large negative $a_0$ signals the existence of a bound state. Taking these values as the approximation of those in the infinite volume, the scattering amplitude ${t}\propto (p\cot\delta_0(p)-ip)^{-1}$ implies a bound state pole at $p_B=i\kappa_B$ with $\kappa_B=49.8_{-18.2}^{+17.0}$ MeV. Consequently, we get the binding energy $E_B$,
\begin{equation}
    E_B=E_{D\bar{D}^*}(p_B)-(m_D+m_{D^*})=-1.3_{-1.0}^{+0.8}~\mathrm{MeV}.
\end{equation}
Weinberg's compositeness criterion states that a near-threshold bound state can be viewed as an admixture of a compact state and a two-hadron channel. According to the generalized Weinberg relation in Ref.~\cite{Li:2021cue}, the small positive $r_0$ indicates the compositeness $X\approx 1$ up to ${O}(p^2)$. Therefore, this bound state is predominantly a $D\bar{D}^*$ bound state.

If the same procedure is applied to cases at other $m_\pi$'s, then similar results can be obtained, as shown in Table~\ref{tab:fitresult}. The ``binding energy" $E_B=-9.7_{-2.2}^{+2.1}~\mathrm{MeV}$ at $m_\pi=250(3)$ MeV is also consistent with the value $11(7)$ MeV obtained in Ref.~\cite{Prelovsek:2013cra} at $m_\pi \approx 266$ MeV. However, the $p^2$ of $E_2$ for these $m_\pi$'s are lower than the corresponding $(p_\mathrm{lhc}^{1\pi})^2$ and make the derived $p\cot\delta_0(p)$ at $E_2$ questionable. Refs.~\cite{Du:2023hlu,Meng:2023bmz} discuss the OPE lhc effect on the $T_{cc}^+(3875)$ relevant $DD^*$ scattering using the lattice data in Ref.~\cite{Padmanath:2022cvl} and find that, except for the singular behavior that $p\cot\delta_0(p)$ bends up abruptly when $p^2$ approaching $(p_\mathrm{lhc}^{1\pi})^2$ from above, the overall $p^2$ behavior is similar to that of ERE including the values of $p^2$ below $(p_\mathrm{lhc}^{1\pi})^2$. Figure~\ref{fig:phase} shows the phase shifts (red points), as well as ERE of $p\cot\delta_0(p)$ (green bands) for the four $m_\pi$'s. In each panel, the ERE curve intersects with the bound state curve $ip$ (dashed green line). If the $D\bar{D}^*$ scattering also has the similar feature to that of the $DD^*(I=0)$ scattering when considering the OPE lhc, then it is possible that $p\cot\delta_0(p)$ has also a pole near the lhc point $(p_\mathrm{lhc}^{1\pi})^2$ (red dashed vertical line) and exhibits a rapid increase there to intersect earlier with the bound state curve $ip$ at $p^2>(p_\mathrm{lhc}^{1\pi})^2$. Therefore, it is very likely that there exists a bound state pole at each $m_\pi$ with a smaller $|E_B|$ than the value in Table~\ref{tab:fitresult}.
\begin{figure}[t]
    \centering
    \includegraphics[width=0.99\linewidth]{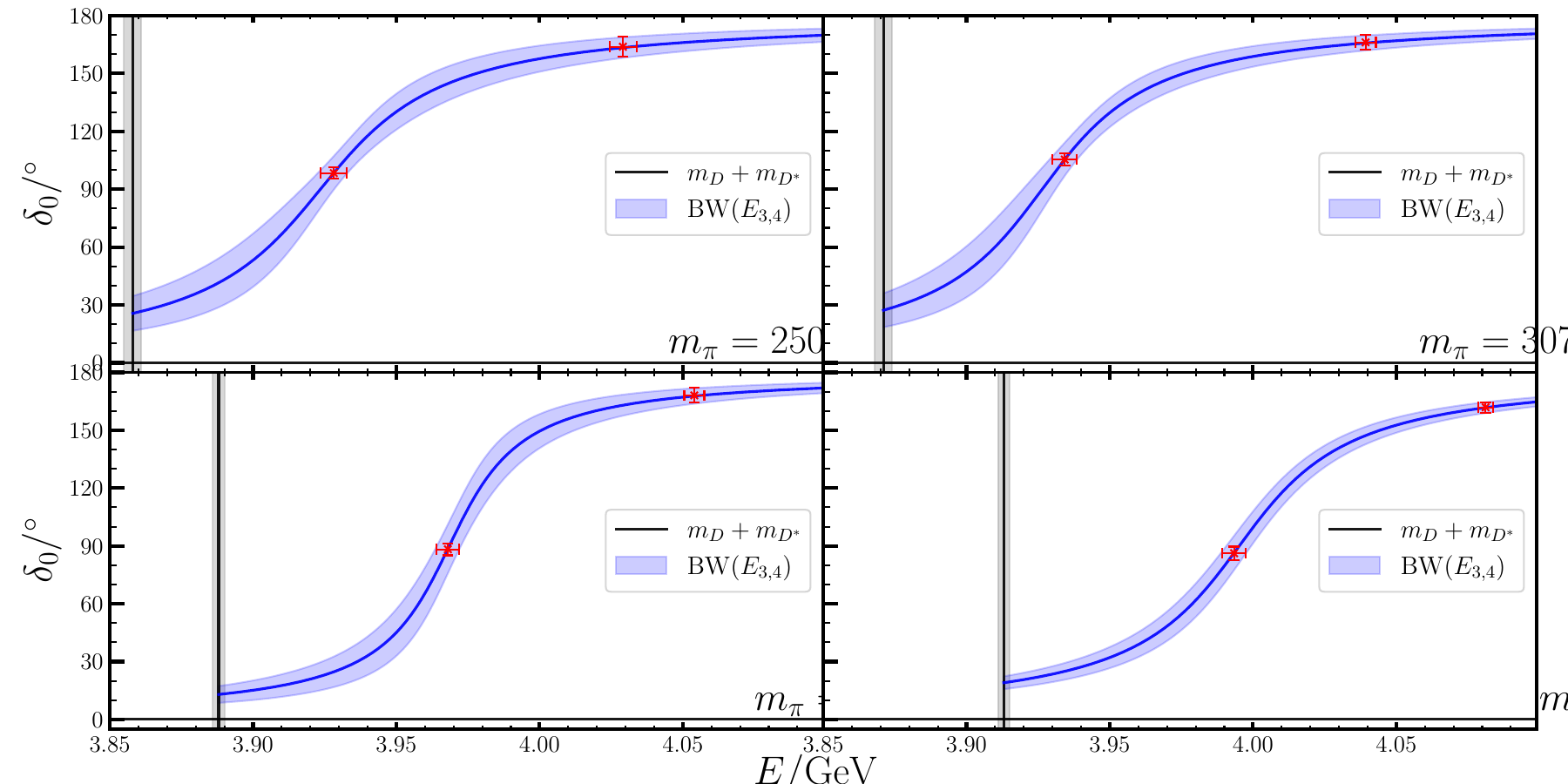}
    \caption{
        The $S$-wave $D\bar{D}^*$ scattering phase $\delta_0$ at $E_{3,4}$ and the Breit-Wigner parameterization. In each panel of the figure, the red data points represent the phase shifts at $E_3$ and $E_4$ at each $m_\pi$ (see Table~\ref{tab:fitresult}). The vertical line indicates the $D\bar{D}^*$ threshold. The blue bands (labelled by BW($E_{2,3}$)) illustrate the resonance using parameters $(m_R,\Gamma_R)$ from $E_{2,3}$ based on the Breit-Wigner ansatz in Eq.~(\ref{eq:resonance}).
    }
    \label{fig:BW}
\end{figure}
\subsection{Possible resonance}
{\it Possible resonance.---}
Now we consider the physical implication of the energy level $E_3$, which lies in the middle of the $D\bar{D}^*$ threshold and $E_{D\bar{D}^*}^{q=1}$. In Ref.~\cite{Prelovsek:2013cra}, the $E_3$ state is interpreted as the upward shifted $D\bar{D}^*$ state with the relative momentum $q=0$ owing to the formation of a bound state below the threshold. The physical consideration behind this is Levinson's theorem that, when $n_l$ bound states are formed in the $l$ partial wave, the phase shift satisfies the relation $\delta_l(p=0^+)=n_l\pi$ with the convention $\delta(p=\infty)=0$. However, when there exist $n_b$ {confined channels (with discretized energies)} between the threshold and a specific momentum $p_\mathrm{max}$, one should use the generalized Levinson's theorem $\delta_l(0^+)-\delta_l(p_\mathrm{max})=(n_l-n_b)\pi$~\cite{Vidal:1992uj,Li:2022aru}. For the case at hand with $n_0=1$ and the possible existence of $\chi_{c1}(2P)$ {(the confined channel) expected by quark models}, $\delta_0(p)$ will undergo an evolution that it starts from $\pi$ at $p=0^+$ ($\cot\delta_0(0^+)=\lim_{p\to 0^+} 1/(a_0 p)\to -\infty$ for $a_0<0$), falls down when the energy increases, and then return to $\pi$ when the energy passes the $\chi_{c1}(2P)$ mass, as is exactly the case (see the values of the phase shift $\delta_0\to 180^\circ$ at $E_4$ in Table~\ref{tab:fitresult}). It is stressed that $E_4>E_{D\bar{D}^*}^{q=1}$ is crucial for this reasoning. Actually, this phase change also hints at the existence of a resonance. The Breit-Wigner ansatz for a resonance $(m_R,\Gamma_R)$ gives
\begin{equation}\label{eq:resonance}
    \delta_0(E)=\arctan\left(\frac{\Gamma_R}{2(m_R-E)}\right).
\end{equation}
Using $\delta_0(E_3)$ and $\delta_0(E_4)$ as inputs, the resonance parameters $(m_R,\Gamma_R)$ at each $m_\pi$ are estimated and are collected in Table~\ref{tab:BW}. Figure~\ref{fig:BW} illustrates the evolution of $\delta_0(E)$ with respect to the center-of-mass energy $E$ of $D\bar{D}^*$, where red points are the lattice data, the blue bands are plots of Eq.~(\ref{eq:resonance}) with parameters in Table~\ref{tab:BW}, and the dashed lines indicate the possible phase change below the resonant energy $E_R$. Since they are tentatively estimated from only two lattice energy levels, the resonance parameters $(m_R,\Gamma_R)$ may change if more scrutinized lattice investigations are performed.

Phenomenological studies also indicate that when the coupled-channel effect of $D\bar{D}^*$ and the quark model $\chi_{c1}(2P)$ is considered, apart from a narrow structure near the $D^0\bar{D}^{*0}$ threshold, there may exist a wide resonance below 4.0 GeV with the width ranging from 17 to 80 MeV~\cite{Kalashnikova:2005ui,Cincioglu:2016fkm,Giacosa:2019zxw,Deng:2023mza,Wang:2023ovj,Wang:2024ytk}.
These observations are compatible with the results of this study. Experimentally,  $X(3940)$ reported by Belle in 2007~~\cite{Belle:2005lik,Belle:2007woe} has the resonance parameters $(m_X,\Gamma)\sim (3942(9), 37_{-17}^{+27})$ MeV, and decays mainly into $D\bar{D}^*$ but not $D\bar{D}$, and therefore favors the assignment of a $1^{++}$ resonance. However, it is puzzling that a similar structure has not been observed by other experiments yet.

\begin{table}[t]
    \caption{Bound state and resonance parameters. The first group of the data lists the resonance parameters $(m_R, \Gamma_R)$ from $E_{3,4}$ are obtained by Eq.~(\ref{eq:resonance}). The second group shows the pole parameters and the pole residuals of the bound state pole through the $K$-matrix paramterization of the phase shifts at $E_{2,3,4}$. The third group gives the resonance parameters as well as the resonance coupling $|c_0|^2$ to $D\bar{D}^*$, the partial decay width $\Gamma_{D\bar{D}^*}$ and the corresponding branching ratio.}
    \label{tab:BW}
    \begin{ruledtabular}
        \begin{tabular}{lrcrr}
            $m_\pi$(MeV)                   & 250(3)      & 307(2)       & 362(1)      & 417(1)      \\\hline
            \multicolumn{5}{c}{BW fit from $E_{3,4}$}                                               \\
            \hline
            $m_R$(MeV)                     & 3924(5)     & 3926(6)      & 3969(4)     & 3995(4)     \\
            $\Gamma_R$(MeV)                & 63(23)      & 57(18)       & 37(13)      & 57(10)      \\\hline
            \multicolumn{5}{c}{{Bound state pole and residual from $E_{2,3,4}$}}                    \\
            \hline
            $\mathrm{Im}p_B$(MeV)          & $+i$ 146(9) & $+i$ 141(10) & $+i$ 57(11) & $+i$ 58(12) \\
            $E_B$(MeV)                     & $-$11(1)    & $-$10(2)     & $-$1.6(7)   & $-$1.7(7)   \\
            $c_0^2$(GeV$^2$)               & 62(4)       & 58(4)        & 23(5)       & 24(5)       \\
            $c_0$(GeV)                     & 7.9(3)      & 7.6(3)       & 4.8(5)      & 4.9(5)      \\
            $\hat{g}$                      & 2.04(7)     & 1.97(8)      & 1.23(12)    & 1.24(14)    \\  
            \hline
            \multicolumn{5}{c}{{Resonance pole and residue from $E_{2,3,4}$}}                       \\
            \hline
            $\mathrm{Re}p_R$(MeV)          & 545(4)      & 559(6)       & 569(4)      & 563(4)      \\
            $\mathrm{Im}p_R$(MeV)          & $-i$ 56(6)  & $-i$  34(8)  & $-i$  38(7) & $-i$  45(7) \\
            $m_R$(MeV)                     & 4008(4)     & 4029(4)      & 4050(3)     & 4071(3)     \\
            $\Gamma_R$(MeV)                & 60(6)       & 38(9)        & 43(8)       & 50(7)       \\
            $|c_0|^2$(GeV$^2$)             & 47(5)       & 28(7)        & 32(6)       & 38(6)       \\
            $|c_0|$(GeV)                   & 6.9(4)      & 5.3(6)       & 5.6(5)      & 6.1(5)      \\
            $\hat{g}$                      & 1.71(9)     & 1.32(17)     & 1.39(13)    & 1.51(12)    \\  
            $\Gamma_{D\bar{D}^*}$(MeV)     & 63(6)       & 39(9)        & 44(8)       & 51(8)       \\
            $\mathrm{Br}_{D\bar{D}^*}(\%)$ & $\sim 100$  & $\sim 100$   & $\sim 100$  & $\sim 100$  \\
        \end{tabular}
    \end{ruledtabular}
\end{table}

\subsection{Joint analysis on $E_{2,3,4}$}
\begin{figure}[t]
    \centering
    \vspace{-0.03in}
    \includegraphics[width=0.98\linewidth]{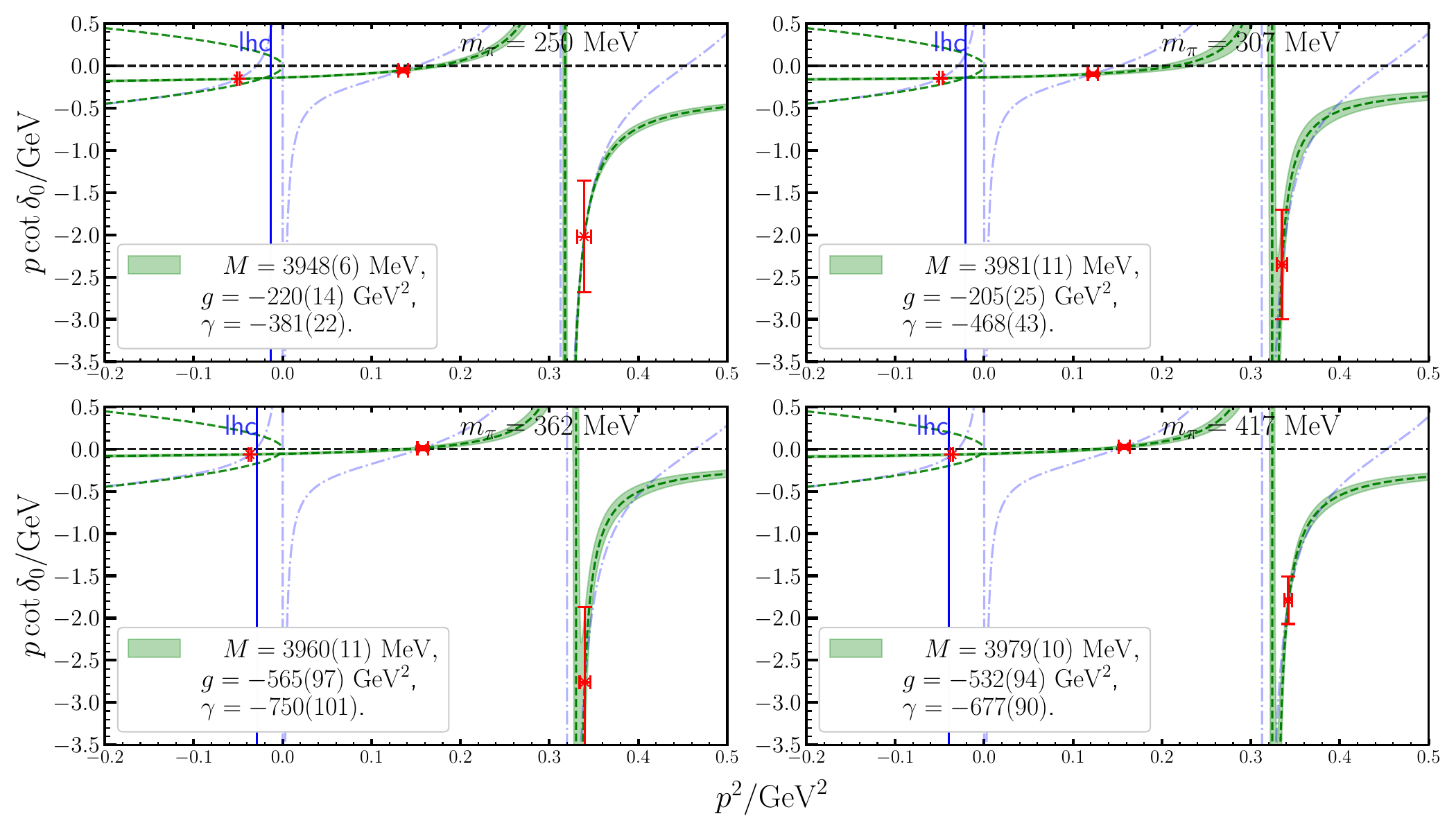}
    \vspace{-0.05in}
    \caption{
    Phase shifts at $E_{2},E_{3}, E_{4}$ and the $K$-matrix parameterization. The layout is similar to that of Fig.~\ref{fig:phase}. The green bands illustrate the $K$-matrix expression of $p\cot\delta_0(p)$ $K$-matrix in Eq.~\ref{eq:reK}. }
    \vspace{-0.15in}
    \label{fig:K-matrix}
\end{figure}

We also perform a joint analysis based on the energy levels $E_{2,3,4}$. The $K$-matrix parameterization is applied to describe the single channel $S$-wave $D\bar{D}^*$ scattering amplitude ${t}$ at $E_{2,3,4}$, namely,
\begin{equation}\label{eq:T-matrix}
    t_0(s) = \frac{K(s)}{1-K(s) i \rho(s)},
\end{equation}
where $s$ is the invariant mass squared of $D\bar{D}^*$, $\rho(s)$ is the two-body phase space factor $\rho(s)=p/(8\pi\sqrt{s})$, and $K(s)$ is a real function of $s$, such that the unitary condition
\begin{equation}\label{eq:T-unitarity}
    \mathrm{Im} {t_0}^{-1}(s) = - \rho(s) = - \frac{1}{16\pi} \frac{2p}{\sqrt{s}} \Theta(\sqrt{s}-E_{\mathrm{thr.}}),
\end{equation}
is satisfied for $s > E_{\mathrm{thr.}}$ with $E_{\mathrm{thr.}}=m_D+m_{D^*}$. In the case of this study, we have the relation
\begin{equation}\label{eq:reK}
    p\cot\delta_0(p(s))= 16\pi \frac{\sqrt{s}}{2} K^{-1}(s).
\end{equation}
As shown in Fig.~\ref{fig:K-matrix}, the phase shift we obtain indicates $p\cot\delta_0(p)\approx 0$ near $p^2(E_3)$, so we parameterize $K(s)$ as
\begin{equation}\label{eq:K-para}
    K(s) = \frac{g}{M^2-s} + \gamma
\end{equation}
with $M,g,\gamma$ being free real parameters, where the single pole term is introduced to account for the possible zero of $K^{-1}(s)$. Thus we tentatively solve the equation given by Eqs.~(\ref{eq:reK}) and (\ref{eq:K-para}) to determine
these parameters using the values of $p\cot\delta_0(p)$ at $E_{2,3,4}$. The obtained parameters $M,g,\gamma$ at the four values of $m_\pi$ are listed in Table~\ref{tab:fitresult}, where the errors are obtained through the jackknife analysis.

The green bands in Fig.~\ref{fig:K-matrix} illustrate the function form in Eq.~(\ref{eq:reK}) with the determined parameters. It is seen that the green bands intersect with the lower branch of the parabola (green dashed line) and indicate the possible existence of a bound state for each $m_\pi$. By solving the pole equation $p\cot\delta_0(p)=ip$ in the physical sheet, we obtain positive imaginary values of $p_B$ and therefore negative binding energies $E_B<0$ for the bound states
at the four $m_\pi$, which are also listed in Table~\ref{tab:fitresult}. The values of $E_B$ are compatible with the results from the $E_{2,3}$ through ERE. The tiny slopes of the green bands near the threshold also indicate small positive effective ranges $r_0$ (if ERE is applicable in this region), which also imply the compositeness $X\approx 1$ of the bound states. Let $s_0=(m_D+m_{D^*}+E_B)^2$ be the bound state pole position in $s$-plane, then close to the pole, the scattering amplitude $t$ has the asymptotic behavior
\begin{equation}\label{eq:pole}
    t_0(s\approx s_0) = \frac{c_0^2}{s_0-s},
\end{equation}
with the pole residual $c_0^2$ giving the coupling of the bound state to $D\bar{D}^*$. With the parameterization of $K(s)$ in Eq.~(\ref{eq:K-para}) and performing a contour integral around $s_0$, we obtain positive values of $c_0^2$, which give the couplings
\begin{eqnarray}\label{eq:X-coupling}
    c_0(m_\pi\approx 250~\mathrm{MeV})&=& 7.9(3)~\mathrm{GeV}\nonumber\\
    c_0(m_\pi\approx 307~\mathrm{MeV})&=& 7.6(3)~\mathrm{GeV}\nonumber\\
    c_0(m_\pi\approx 362~\mathrm{MeV})&=& 4.8(5)~\mathrm{GeV}\nonumber\\
    c_0(m_\pi\approx 417~\mathrm{MeV})&=& 4.9(5)~\mathrm{GeV}.
\end{eqnarray}
By assuming a pure bound state for $X(3872)$ (the compositeness $X\approx 1$), an effective field theory study comes out with the effective coupling~\cite{Guo:2013zbw}
\begin{equation}
    c_0^2\approx \frac{16\pi}{\mu}\sqrt{\frac{2|E_B|}{\mu}}\big(m_D m_{D^*} \sqrt{s_0}\big),
\end{equation}
with $\mu=(m_D+m_{D^*})/(m_D m_{D^*})$ being the reduced mass of the $D\bar{D}^*$ system, which gives $c_0\approx(10.6_{-0.3}^{+0.3},10.4_{-0.6}^{+0.5},6.7_{-0.8}^{+0.6}, 6.7_{-0.8}^{+0.6})~\mathrm{GeV}$ at the four pion masses $m_\pi=(250,307,362,417)~\mathrm{MeV}$. These values are qualitatively compatible with those in Eq.~(\ref{eq:X-coupling}) from the pole residuals. In addition, these bound state poles also pass the sanity check that the $S$-matrix satisfies the asymptotic behavior $S(p\approx p_B)\approx \frac{-ig_B^2}{p-p_B}$ with $g_B^2=c_0^2\mu/ \sqrt{s_0}>0$~\cite{Iritani:2017rlk}.

\begin{figure}[t]
    \centering
    \vspace{-0.03in}
    \includegraphics[width=0.48\linewidth]{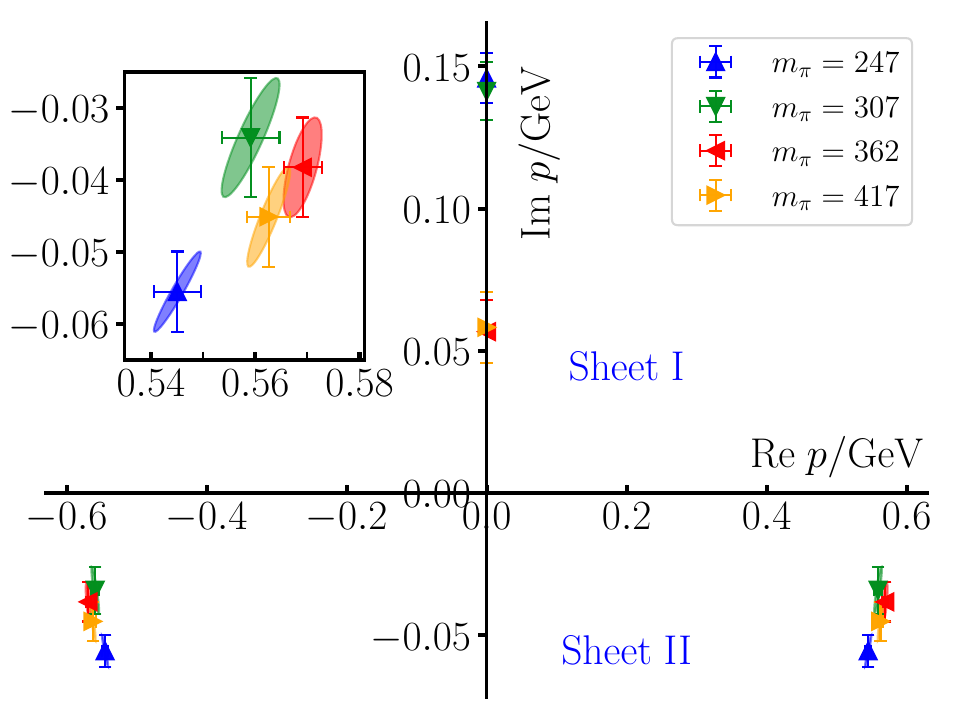}
    \includegraphics[width=0.48\linewidth]{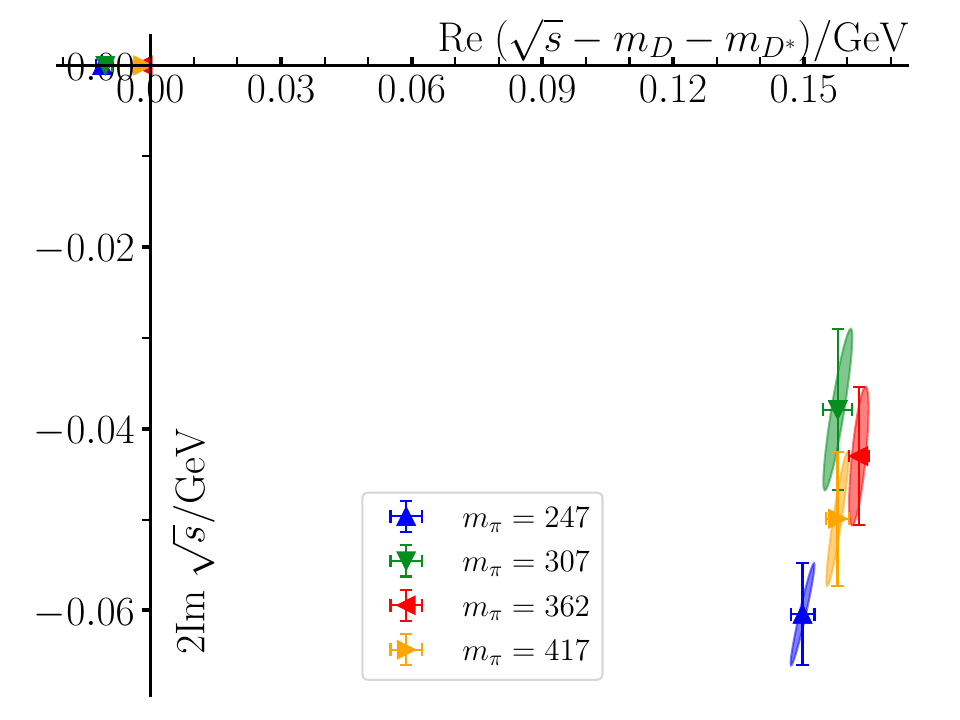}
    \vspace{-0.05in}
    \caption{
        Pole positions in the complex $p$-plane (left) and $\sqrt{s}$-plane (right). Only physical poles are shown. The bound state poles reside on the positive imaginary axis of the $p$-plane (left) and the negative real axis of $\sqrt{s}$-plane (right). The points in the lower half of the $p$-plane (left) and the sight-bottom $\sqrt{s}$-plane illustrate the resonance poles.}
    \vspace{-0.15in}
    \label{fig:pole-position}
\end{figure}

{
With the parameters $M,g,\gamma$, we check if there exist resonance poles by solving the pole equation $p(s)\cot\delta_0(p(s))-ip(s)=0$ in the unphysical Riemann sheet with $\mathrm{Im}p(s)<0$. A pair of conjugate resonance poles $\sqrt{s_0}=m_R\pm i \frac{\Gamma_R}{2}$ are obtained for each $m_\pi$,
\begin{eqnarray}\label{eq:Resonance-pole}
    (m_R,\Gamma_R)(m_\pi\approx 250~\mathrm{MeV})&=& (4008(4),60(6))~\mathrm{MeV}\nonumber\\
    (m_R,\Gamma_R)(m_\pi\approx 306~\mathrm{MeV})&=& (4029(4),38(9))~\mathrm{MeV}\nonumber\\
    (m_R,\Gamma_R)(m_\pi\approx 362~\mathrm{MeV})&=& (4050(3),43(8))~\mathrm{MeV}\nonumber\\
    (m_R,\Gamma_R)(m_\pi\approx 417~\mathrm{MeV})&=& (4071(3),50(7))~\mathrm{MeV},\nonumber\\
\end{eqnarray}
whose positions in the complex $p$-plane and $\sqrt{s}$-plane are illustrated in Fig.~\ref{fig:pole-position}. We also use Eq.~(\ref{eq:pole}) to extract the resonance coupling $c_0$ at the pole $s\approx s_0$ and obtain complex values of $c_0^2$ at different $m_\pi$. Their modulus $|c_0|^2, |c_0|$ and the corresponding dimensionless coupling $\hat{g}=|c_0|/m_R$ are collected in Table~\ref{tab:BW}. With the coupling $|c_0|^2$ of the resonance $R$ to $D\bar{D}^*$, the partial decay width $\Gamma_{D\bar{D}^*}$ is predicted using
\begin{equation}
    \Gamma_{D\bar{D}^*}=|c_0|^2 \frac{\rho(m_R^2)}{m_R},
\end{equation}
which are also listed in Table~\ref{tab:BW}. It is seen that, for all the four $m_\pi$, we have $\Gamma_{D\bar{D}^*}\approx \Gamma_R$, which implies that the decay branching fraction $\Gamma_{D\bar{D}^*}/\Gamma_R$ is almost 100\%, as is expected since $D\bar{D}^*$ is the only open-charm decay channel permitted by the kinematics and symmetries.}

\begin{figure}[t]
    \centering
    \vspace{-0.03in}
    \includegraphics[width=0.99\linewidth]{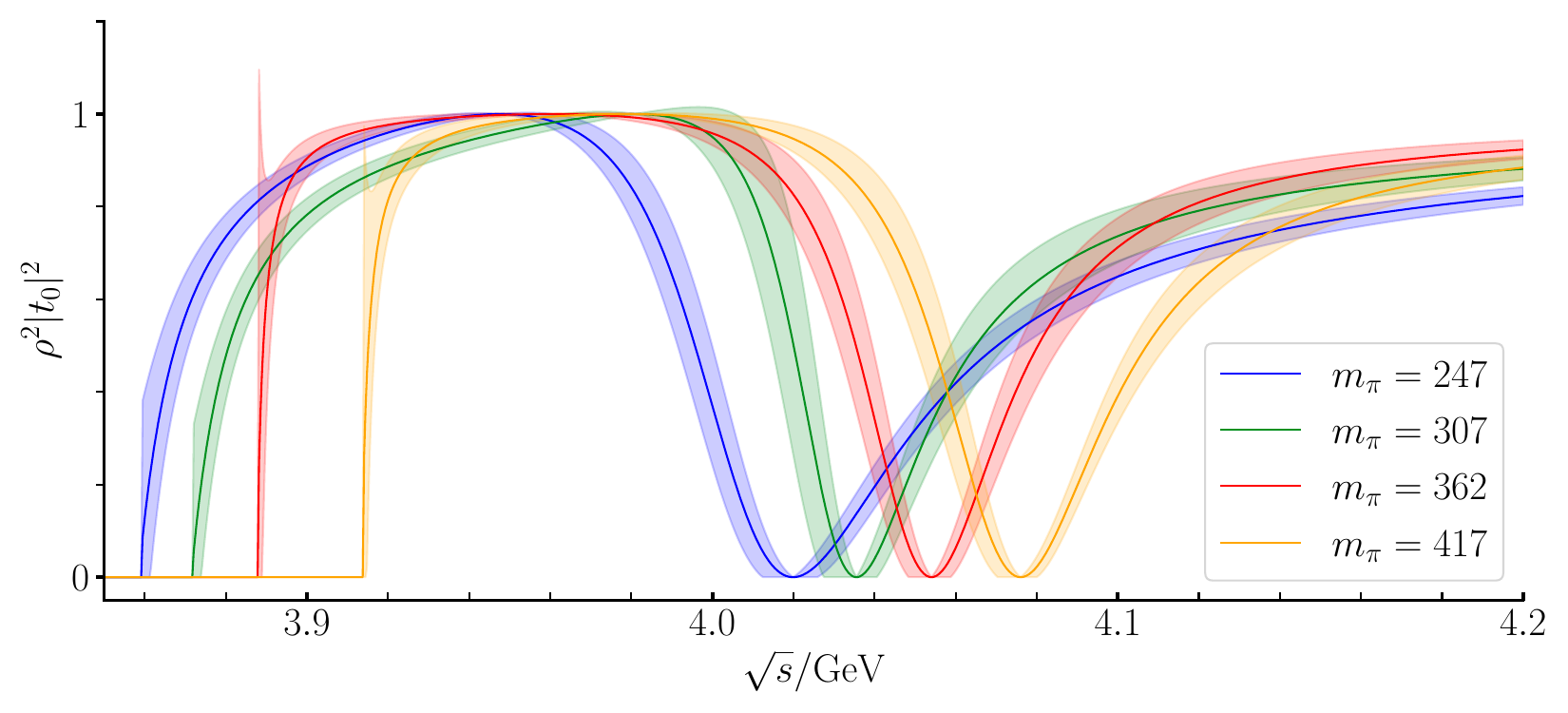}
    \vspace{-0.05in}
    \caption{Amplitudes $\rho^2 |t_0|^2$ for the four pion masses. The zero values of $\rho^2|t_0|^2$ signal the presence of the CDD zero~\cite{Castillejo:1955ed} due to the existence of additional components (the quark model $\chi_{c1}(2P)$ for example).
    }
    \vspace{-0.15in}
    \label{fig:cross-section}
\end{figure}

We also plot in Fig.~\ref{fig:cross-section} the squared $D\bar{D}^*\to D\bar{D}^*$ scattering amplitude $\rho^2 |t_0|^2$ for $\sqrt{s}>E_\mathrm{thr.}$. For each $m_\pi$, the zero value of $\rho^2|t_0|^2$ is due to the well-known Castillejo-Dalitz-Dyson (CDD) zero~\cite{Castillejo:1955ed} of the amplitude $t_0$, which signals the possible existence of an additional component other than the $D\bar{D}^*$ system (the quark model $\chi_{c1}(2P)$ for instance)~\cite{Chew:1961cer}. We notice that a model study on the two-particle scattering shows the presence of the CDD zero when there exists a bare state in the relevant energy region~\cite{Li:2022aru}. Theoretically, the vanishing $t_0$ at the CDD zero can be attributed to the cancellation of the $D\bar{D}^*$ self-interaction and the $D\bar{D}^*$-$\chi_{c1}(2P)$ interaction owing to the interplay of hadron and quark degrees of freedom~\cite{Hyodo:2008xr,Baru:2010ww,Hanhart:2011jz,Kamiya:2016oao,Guo:2016wpy,Kang:2016jxw}. The positions $m_\mathrm{CDD}$ of the CDD zeros for the four values of $m_\pi$ are determined to be

\begin{eqnarray}
    m_\mathrm{CDD}(m_\pi\approx 250~\mathrm{MeV})&=& 4020(3)~\mathrm{MeV}\nonumber\\
    m_\mathrm{CDD}(m_\pi\approx 306~\mathrm{MeV})&=& 4036(3)~\mathrm{MeV}\nonumber\\
    m_\mathrm{CDD}(m_\pi\approx 362~\mathrm{MeV})&=& 4054(3)~\mathrm{MeV}\nonumber\\
    m_\mathrm{CDD}(m_\pi\approx 417~\mathrm{MeV})&=& 4076(2)~\mathrm{MeV},
\end{eqnarray}
which are very close to the corresponding resonance masses $m_R$ and imply a close connection of the resonance with the possible quark model $\chi_{c1}(2P)$.

The LHCb collaboration recently reported the clear observation of a $1^{++}$ charmoniumlike state $\chi_{c1}(4010)$ with the resonance parameters $(m_{\chi_{c1}},\Gamma_{\chi{c1}}=(4012.5_{-3.9-3.7}^{+3.6+4.1},62.7_{-6.4-6.6}^{+7.0+6.4})~\mathrm{MeV}$~\cite{LHCb:2024vfz}. It is amazing that the resonance parameters we obtain in Eq.~(\ref{eq:Resonance-pole}) are in quite good agreement with $\chi_{c1}(4010)$. We are not sure at present whether this coincidence of the lattice QCD result and the experiment is accidental or a glimpse of the truth. After all, we have only three finite volume energy levels $E_{2,3,4}$ at hand, it is not the time to draw a sound conclusion until more scrutinized lattice QCD studies on this topic are performed in the future.

\section{Summary}
We calculate the $I^G J^{PC}=0^+1^{++}$ channel $D\bar{D}^*$ scattering at four pion masses ranging from 250 MeV to 417 MeV in $N_f=2$ lattice QCD. We use the distillation method to calculate the related correlation matrix of $c\bar{c}$ operators and $D\bar{D}^*$ operators and obtain three energy $E_{2,3,4}$ levels around the $D\bar{D}^*$ threshold. At $m_\pi\approx 417$ MeV where there are no severe OPE lhc issues, a normal ERE analysis to the two energy levels $E_{2,3}$ results in a shallow bound state, which resides below the $D\bar{D}^*$ by $1.3_{-1.0}^{+0.8}$ MeV and has a compositeness $X\approx 1$ indicating a predominant $D\bar{D}^*$ component. The phase shifts $p\cot\delta_0(p)$ at $E_{3,4}$ indicate the possible existence of a resonance below $4.0$ GeV.

We also perform a joint analysis of the three energy levels $E_{2,3,4}$ through the $K$-matrix parameterization of the scattering amplitude. After solving the pole equation in the physical Riemann sheet, a genuine bound state is obtained, whose properties are similar to those of the bound state derived through the ERE analysis.
This bound state can correspond to $X(3872)$. In addition, we observe a resonance pole in the second Riemann sheet with a mass slightly higher than 4.0 GeV and a width of 40-60 MeV. This resonance decays into $D\bar{D}^*$ by almost 100\%. It is interesting to note that the properties of this resonance are in fairly good agreement with the newly observed $\chi_{c1}(4010)$ by LHCb~\cite{LHCb:2024vfz}. Anyway, we remark that this result should be checked by more sophisticated lattice QCD calculations in the future with more finite volume energy levels from more lattice volumes and more kinetic frames~\cite{Prelovsek:2020eiw,Wilson:2023anv,Wilson:2023hzu}.

\vspace{0.5cm}
\begin{acknowledgments}
    We are grateful to Profs. Qiang Zhao, Feng-Kun Guo, Jia-Jun Wu, Zhihui Guo and Xu Feng for the valuable discussions.
    This work is supported by the National Natural Science Foundation of China (NNSFC) under Grants No. 11935017, No. 12293060, No. 12293061, No. 12293065, No. 12293063, No. 12075253, No. 12070131001 (CRC 110 by DFG and NNSFC), No. 12175063, No. 12205311, the National Key Research and Development Program of China (No. 2020YFA0406400) and the Strategic Priority Research Program of Chinese Academy of Sciences (No. XDB34030302). The Chroma software system~\cite{Edwards:2004sx} and QUDA library~\cite{Clark:2009wm,Babich:2011np} are acknowledged. The computations were performed on the HPC clusters at the Institute of High Energy Physics (Beijing), China Spallation Neutron Source (Dongguan), and the ORISE computing environment.
\end{acknowledgments}

\input{appendix}

\clearpage
\bibliography{ref}

\end{document}

%% file: appendix.tex








\setcounter{figure}{0}
\renewcommand{\thefigure}{A\arabic{figure}}
\setcounter{table}{0}
\renewcommand{\thetable}{A\arabic{table}}

\appendix
\section{Conventions and Operators}\label{sec:appendix-S1}
These appendix provide further information on our study on the $D\bar{D}^*(I=0)$ scattering process.
The contents include the details of the operators' construction, the calculation of correlation matrix, and the derivation of finite volume energy levels, etc.

We take the following procedure to build $D\bar{D}^*$ operators with quantum numbers $I^{G} J^{PC} = 0^+1^{++}$. According to the isospin assignment of $u,d$ quarks,
\begin{equation}
    \begin{cases}
        |u\rangle=\left|\frac{1}{2}, \frac{1}{2}\right\rangle, \\

        |d\rangle=\left|\frac{1}{2},-\frac{1}{2}\right\rangle,
    \end{cases}
    \begin{cases}
        |\bar{d}\rangle=\left|\frac{1}{2}, \frac{1}{2}\right\rangle, \\
        |\bar{u}\rangle=-\left|\frac{1}{2},-\frac{1}{2}\right\rangle ,
    \end{cases}
\end{equation}
we have the isospin states of $D$ and $D^*$,
\begin{equation}
    \begin{cases}
        |D^{(*)+}\rangle = |c\bar{d}\rangle = |\frac{1}{2},\frac{1}{2}\rangle ,       \\
        |D^{(*)0}\rangle = |c\bar{u}\rangle = -|\frac{1}{2},-\frac{1}{2}\rangle ,     \\
        |\bar{D}^{(*)0}\rangle = |u\bar{c}\rangle = |\frac{1}{2},\frac{1}{2}\rangle , \\
        |D^{(*)-}\rangle = |d\bar{c}\rangle = |\frac{1}{2},-\frac{1}{2}\rangle.       \\
    \end{cases}
    \label{eq:D-meson-wave-function}
\end{equation}
Thus the $I=0$ combination of charge neutral $D\bar{D}^*$ states reads
\begin{eqnarray}\label{eq:SM-isospin}
    |D\bar{D}^*\rangle_{I=0}^{Q=0}&=&\frac{1}{2}\left(|D^+\bar{D}^{*-}\rangle+|D^0\bar{D}^{*0}\rangle \right.\nonumber\\
                                  &&\left.-|\bar{D}^0 D^{*0}\rangle-|D^-D^{*+}\rangle\right).
\end{eqnarray}
Since the charge conjugation transformation ($\mathcal{C})$ of $D$ and $D^*$ are conventionally defined as $\mathcal{C} | D \rangle = + |\bar{D}\rangle$, $\mathcal{C} | D^* \rangle = - |\bar{D}^* \rangle$, one can easily check the $\mathcal{C}$ transformation $\mathcal{C}|D\bar{D}^*\rangle_{I=0}^{Q=0}=+|D\bar{D}^*\rangle_{I=0}^{Q=0}$. The $\mathcal{P}$-parity $P=+$ requires the $D\bar{D}^*$ relative angular momentum $l=0$. Therefore, the $S$-wave $D\bar{D}^*$ state with the combination in Eq.~(\ref{eq:SM-isospin}) has the desired quantum numbers $I^{G} J^{PC} = 0^+1^{++}$. The combination in Eq.~(\ref{eq:SM-isospin}) gives also the flavor structure of the interpolation
field operator of $D\bar{D}^*$ systems.

A $S$-wave two-particle operator $\mathcal{O}_{AB}(q)$ is easily constructed by summing over all the relative momenta $\mathbf{p}\equiv 2\pi \mathbf{q}/L $ with the same $q=|\mathbf{q}|$, namely,
\begin{equation}\label{eq:SM-swave}
    \mathcal{O}_{AB}^{q} = \frac{1}{N_q}\sum\limits_{R \in O} \mathcal{O}_{A}(R \circ \mathbf{q}) \mathcal{O}_{B}(- R \circ \mathbf{q}).
\end{equation}
where $R$ runs over all the elements of the octahedral group $O$ and $N_q$ is the degeneracy of $q$. Given the single-particle operators of $D$ and $\bar{D}^*$, $(\mathcal{O}_D, \mathcal{O}_{D^*})\equiv (\bar{u}(\bar{d})\Gamma_D c,\bar{u}(\bar{d})\Gamma_{D^*} c)$, we combine the relations in Eq.~(\ref{eq:SM-isospin}) and (\ref{eq:SM-swave}) to build up the $D\bar{D}^*$ operators $\mathcal{O}_{D\bar{D}^*}^{q}$ with the quantum numbers $I^{G} J^{PC} = 0^+1^{++}$. In practice, we use two combinations of $\Gamma_{D}$ and $\Gamma_{D^*}$, namely,  $(\Gamma_D,\Gamma_{D^*})=(\gamma_5,\gamma_i)$ and $(\gamma_4\gamma_5,\gamma_4\gamma_i)$. The operators $\mathcal{O}_{D\bar{D}^*}^{q=0,\gamma_5(\gamma_4\gamma_5)}$ with the two $(\Gamma_D,\Gamma_{D^*})$ combinations (indicated by the superscripts $\gamma_5$ and $\gamma_4\gamma_5$) are also named by $\mathcal{O}_4$ and $\mathcal{O}_5$, respectively. The $\mathcal{O}_{D\bar{D}^*}^{q=1}$ operator uses the first $(\Gamma_D,\Gamma_{D^*})$ combination and is also named as $\mathcal{O}_6$.

The $S$-wave $J/\psi\omega$ operator with $q=0$ is simply the antisymmetric combination of the single particle operators $\mathcal{O}_{J/\psi}^j(q=0)$ and $\mathcal{O}_\omega^k (q=0)$, namely, $\mathcal{O}_7\equiv\mathcal{O}_{J/\psi\omega}^{q=0}\to \epsilon_{ijk}\mathcal{O}_{J/\psi}^j(q=0)\mathcal{O}_\omega^k(q=0)$.

For the convenience of discussions, we recite the build-up of $c\bar{c}$ operators presented in the main text. Since the $\chi_{c1}$ mass $m_{\chi_{c1}}\approx 3.51~\mathrm{GeV}$ is far from the energy region relevant to $X(3872)$, we need several $c\bar{c}$ operators to distinguish states of interest from $\chi_{c1}$. So we build several spatially extended charmonium $c\bar{c}$ operators
\begin{equation}
    \mathcal{O}_{c\bar{c}}^r(t)=\frac{1}{N_r}\sum\limits_{|\mathbf{y-x}|=r}\bar{c}(\mathbf{x},t) \gamma_5\gamma_i K_U (\mathbf{x},\mathbf{y};t)c(\mathbf{y},t),
\end{equation}
where $N_r$ is the multiplicity of $\mathbf{r}=\mathbf{y-x}$ with $|\mathbf{r}|=r$, and
\begin{equation}
    K_U(\mathbf{x},\mathbf{y};t)=\mathcal{P}e^{ig\int_\mathbf{y}^\mathbf{x}\mathbf{A}\cdot d\mathbf{r}}
\end{equation}
is a Wilson line connecting $(\mathbf{y},t)$ and $(\mathbf{x},t)$. Obviously, $\mathcal{O}_{c\bar{c}}^r(t)$ is gauge invariant and has the right quantum number $J^{PC}=1^{++}$ (actually $T_1^{++}$ on the lattice) after the summation over $|\mathbf{r}|=r$. In practice, we use three $\mathcal{O}_{c\bar{c}}^r(t)$ operators (denoted by $\mathcal{O}_{1,2,3}$) with $r/a_s=0,1,2$, respectively.

\begin{figure*}[t]
    \centering
    \includegraphics[width=1\linewidth]{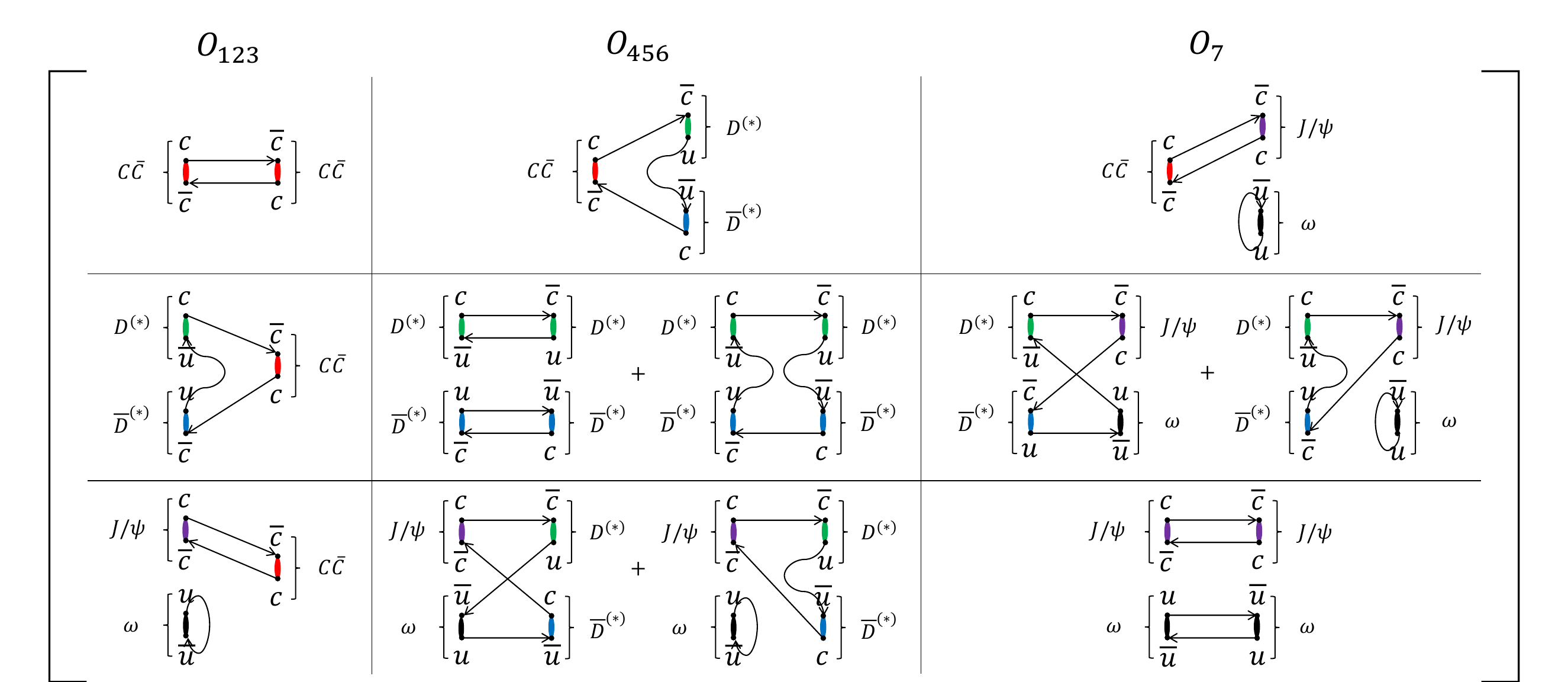}
    \caption{Schematic quark contraction diagrams for the calculation of the correlation matrix $C_{\alpha\beta}(t)$.
    For convenience, the operator set is sorted as $\{\mathcal{O}_{\alpha}| \alpha = 1,\cdots, 7\} = \{ \mathcal{O}_{c\bar{c}}^{r=0}, \mathcal{O}_{c\bar{c}}^{r=1}, \mathcal{O}_{c\bar{c}}^{r=2}, \mathcal{O}_{D\bar{D}}^{\mathbf{q}=0, \gamma_5}, \mathcal{O}_{D\bar{D}}^{{q}=0, \gamma_4\gamma_5}, \mathcal{O}_{D\bar{D}}^{{q}=1}, \mathcal{O}_{J/\psi\omega}^{{q}=0}\}$.
    All the contributions are considered except for charm quark annihilation diagrams which are expected to be suppressed by the OZI rule.
    }
    \label{fig:SM_schematic-diagram}
\end{figure*}

\begin{figure*}[t]
    \includegraphics[width=0.49\linewidth]{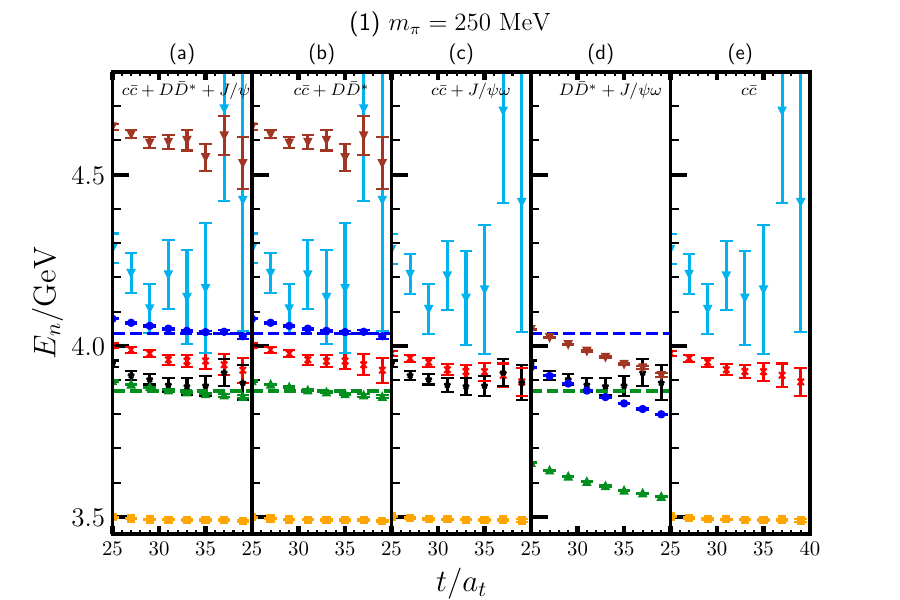}
    \includegraphics[width=0.49\linewidth]{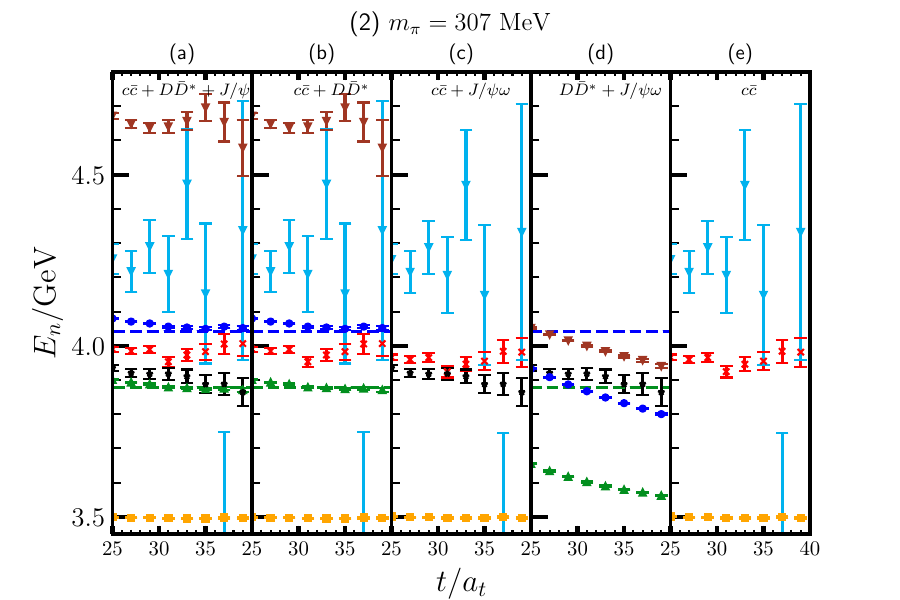}
    \includegraphics[width=0.49\linewidth]{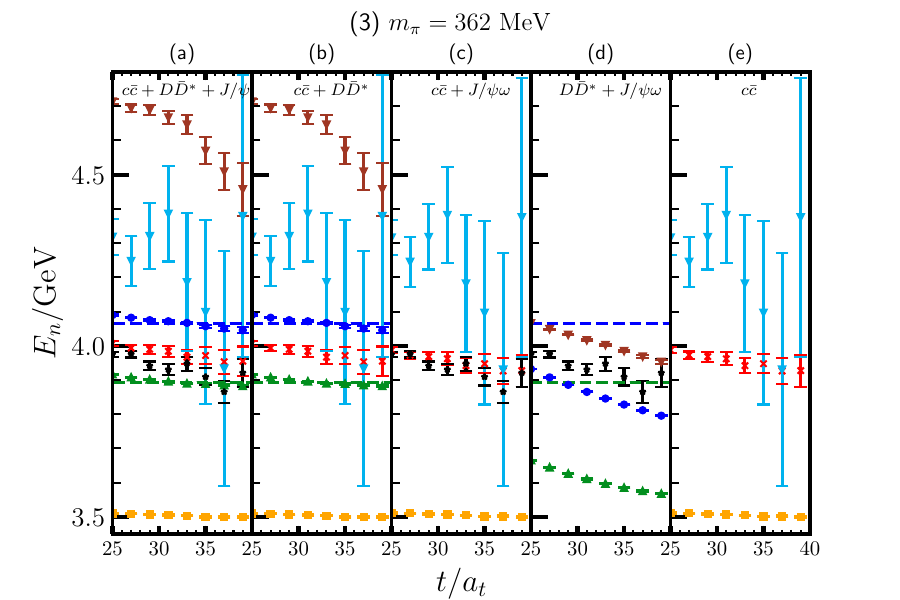}
    \includegraphics[width=0.49\linewidth]{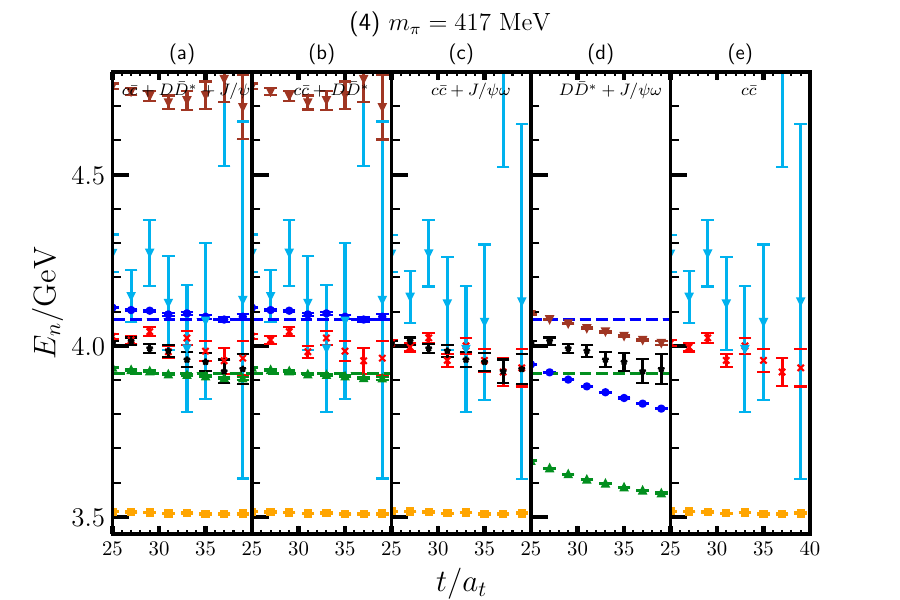}
    \caption{Effective mass using different operators subset and solving GEVP, ranging from different $m_\pi$. In each column of the panels, the headers indicate the operators involved in GEVP. The colored lines show the effective mass behavior corresponding to their dominant operators, as indicated by the same color. Their fitting values are given in Table.[\ref{tab:energies-M245}-\ref{tab:energies-M415}].}
    \label{fig:SM_effmass_gevp}
\end{figure*}
\begin{table*}[t]
    \caption{Energy levels $E_n$ from different subsets of operators for lattice ensemble $m_\pi=250~\mathrm{MeV}$. In each row of $E_n$, the energy values (converted to the physical unit GeV through the fixed $a_t^{-1}$ in Table~\ref{tab:SM-anisotropy-parameter}) are
    obtained from the operators in the non-empty columns. Therefore, the first four rows of $E_n$ show the unmixed energies. The mixing effects are very weak and can almost be neglected for $D\bar{D}^*(q=0)-D\bar{D}^*(q=1)$ (row no.5), $J/\psi\omega-D\bar{D}^*$ (row no.6) and $J/\psi\omega-c\bar{c}$ (row no.7) mixings. The strong mixing takes place between $c\bar{c}$ states and $D\bar{D}^*$ states, which is manifested by the substantial changes of energies $E_2$ and $E_3$ after the mixing (row no.8 and row no.9).}
    \label{tab:energies-M245}
    \begin{ruledtabular}
        \begin{tabular}{c|ccc|ccc|c}
            $\mathcal{O}_\alpha$ &            & $\mathcal{O}_{1,2,3}$ &           & $\mathcal{O}_4(D\bar{D}^*,q=0)$ & $\mathcal{O}_5(D\bar{D}^*,q=0)$ & $\mathcal{O}_6(D\bar{D}^*,q=1)$ & $\mathcal{O}_7(J/\psi\omega,q=0)$ \\ \hline
            $n$                  & 1          & 3                     & 5         & 2                               & 4                               & 6                               & -                                 \\ \hline
            $E_n$(GeV)           & 3.4907(20) & {\bf 3.9082(47)}      & 4.087(54) &                                 &                                 &                                 &                                   \\
                                 & 3.4907(20) & {\bf 3.9082(47)}      & 4.087(54) &                                 &                                 &                                 & 3.889(13)                         \\
                                 &            &                       &           &                                 &                                 &                                 &                                   \\
                                 &            &                       &           & 3.5099(65)                      & 4.142(22)                       & 4.215(49)                       &                                   \\
                                 &            &                       &           & 3.5099(65)                      & 4.142(22)                       & 4.215(49)                       & 3.889(13)                         \\
                                 &            &                       &           &                                 &                                 &                                 &                                   \\
                                 & 3.4896(19) & {\bf 3.9282(46)}      & 4.089(55) & {\bf 3.8324(36)}                & \bf{4.0308(49)}                 & 4.577(14)                       & 3.889(13)                         \\
                                 & 3.4896(19) & {\bf 3.9282(46)}      & 4.089(55) & {\bf 3.8324(36)}                & \bf{4.0308(49)}                 & 4.577(14)                       &
        \end{tabular}
    \end{ruledtabular}
\end{table*}
\begin{table*}[t]
    \caption{Similar to Tab.~\ref{tab:energies-M245} but for $m_\pi=307~\mathrm{MeV}$.}
    \label{tab:energies-M305}
    \begin{ruledtabular}
        \begin{tabular}{c|ccc|ccc|c}
            $\mathcal{O}_\alpha$ &            & $\mathcal{O}(r=0,a_s,2a_s)$ &           & $\mathcal{O}_{D\bar{D}^*}^{1}(q=0)$ & $\mathcal{O}_{D\bar{D}^*}^{2}(q=0)$ & $\mathcal{O}_{D\bar{D}^*}^{1}(q=1)$ & $J/\psi\omega$ \\ \hline
            $n$                  & 1          & 3                           & 5         & 2                                   & 4                                   & 6                                   & -             \\ \hline
            $E_n$(GeV)           & 3.4947(20) & {\bf 3.9135(47)}            & 4.178(45) &                                     &                                     &                                     &               \\
                                 & 3.4947(20) & {\bf 3.9135(47) }            & 4.178(45) &                                     &                                     &                                     & 3.878(13)     \\
                                 &            &                             &           &                                     &                                     &                                     &               \\
                                 &            &                             &           & 3.5180(67)                          & 4.135(24)                           & 4.133(47)                           &               \\
                                 &            &                             &           & 3.5180(67)                          & 4.135(24)                           & 4.133(47)                           & 3.878(13)     \\
                                 &            &                             &           &                                     &                                     &                                     &               \\
                                 & 3.4935(19) & {\bf 3.9343(43)}            & 4.182(45) & {\bf 3.8461(33)}                    & {\bf 4.0411(36)}                    & 4.619(13)                           & 3.878(13)     \\
                                 & 3.4935(19) & {\bf 3.9343(43)}            & 4.182(45) & {\bf 3.8461(33)}                    & {\bf 4.0411(36)}                    & 4.619(13)                           &
        \end{tabular}
    \end{ruledtabular}
\end{table*}
\begin{table*}[t]
    \caption{Similar to Tab.~\ref{tab:energies-M245} but for $m_\pi=362~\mathrm{MeV}$.}
    \label{tab:energies-M365}
    \begin{ruledtabular}
        \begin{tabular}{c|ccc|ccc|c}
            $\mathcal{O}_\alpha$ &            & $\mathcal{O}(r=0,a_s,2a_s)$ &           & $\mathcal{O}_{D\bar{D}^*}^{1}(q=0)$ & $\mathcal{O}_{D\bar{D}^*}^{2}(q=0)$ & $\mathcal{O}_{D\bar{D}^*}^{1}(q=1)$ & $J/\psi\omega$ \\ \hline
            $n$                  & 1          & 3                           & 5         & 2                                   & 4                                   & 6                                   & -             \\ \hline
            $E_n$(GeV)           & 3.5021(23) & {\bf 3.9528(44)}            & 4.242(41) &                                     &                                     &                                     &               \\
                                 & 3.5021(23) & {\bf 3.9528(44)}            & 4.242(41) &                                     &                                     &                                     & 3.923(10)     \\
                                 &            &                             &           &                                     &                                     &                                     &               \\
                                 &            &                             &           & 3.5208(61)                          & 4.131(21)                           & 4.187(37)                           &               \\
                                 &            &                             &           & 3.5208(61)                          & 4.131(21)                           & 4.187(37)                           & 3.923(10)     \\
                                 &            &                             &           &                                     &                                     &                                     &               \\
                                 & 3.5018(21) & {\bf 3.9680(40)}            & 4.245(41) & {\bf 3.8687(26)}                    & {\bf 4.0589(36)}                    & 4.688(11)                           & 3.923(10)     \\
                                 & 3.5018(21) & {\bf 3.9680(40)}            & 4.245(41) & {\bf 3.8687(26)}                    & {\bf 4.0589(36)}                    & 4.688(11)                           &
        \end{tabular}
    \end{ruledtabular}

\end{table*}
\begin{table*}[t]
    \caption{Similar to Tab.~\ref{tab:energies-M245} but for $m_\pi=417~\mathrm{MeV}$.}
    \label{tab:energies-M415}
    \begin{ruledtabular}
        \begin{tabular}{c|ccc|ccc|c}
            $\mathcal{O}_\alpha$ &            & $\mathcal{O}(r=0,a_s,2a_s)$ &           & $\mathcal{O}_{D\bar{D}^*}^{1}(q=0)$ & $\mathcal{O}_{D\bar{D}^*}^{2}(q=0)$ & $\mathcal{O}_{D\bar{D}^*}^{1}(q=1)$ & $J/\psi\omega$ \\ \hline
            $n$                  & 1          & 3                           & 5         & 2                                   & 4                                   & 6                                   & -             \\ \hline
            $E_n$(GeV)           & 3.5089(21) & {\bf 3.9778(43)}            & 4.121(73) &                                     &                                     &                                     &               \\
                                 & 3.5089(21) & {\bf 3.9778(43)}            & 4.121(73) &                                     &                                     &                                     & 3.9688(77)    \\
                                 &            &                             &           &                                     &                                     &                                     &               \\
                                 &            &                             &           & 3.5320(55)                          & 4.179(24)                           & 4.348(70)                           &               \\
                                 &            &                             &           & 3.5320(55)                          & 4.179(24)                           & 4.348(70)                           & 3.9688(77)    \\
                                 &            &                             &           &                                     &                                     &                                     &               \\
                                 & 3.5083(20) & {\bf 3.9934(41)}            & 4.123(74) & {\bf 3.8949(24)}                    & {\bf 4.0850(27)}                    & 4.720(11)                           & 3.9688(77)    \\
                                 & 3.5083(20) & {\bf 3.9934(41)}            & 4.123(74) & {\bf 3.8949(24)}                    & {\bf 4.0850(27)}                    & 4.720(11)                           &
        \end{tabular}
    \end{ruledtabular}
\end{table*}

\setcounter{figure}{0}
\renewcommand{\thefigure}{B\arabic{figure}}
\setcounter{table}{0}
\renewcommand{\thetable}{B\arabic{table}}
\section{Determination of finite volume energies}\label{sec:appendix-S2}

\begin{widetext}
Based on the operator set introduced above,
\begin{equation}
    \mathcal{S}=\{\mathcal{O}_{\alpha}| \alpha = 1,\cdots, 7\} = \{ \mathcal{O}_{c\bar{c}}^{r=0}, \mathcal{O}_{c\bar{c}}^{r=1}, \mathcal{O}_{c\bar{c}}^{r=2}, \mathcal{O}_{D\bar{D}}^{\mathbf{q}=0, \gamma_5}, \mathcal{O}_{D\bar{D}}^{{q}=0, \gamma_4\gamma_5}, \mathcal{O}_{D\bar{D}}^{{q}=1}, \mathcal{O}_{J/\psi\omega}^{{q}=0}\},
\end{equation}
we calculate the corresponding correlation matrix
\begin{equation}
    C_{\alpha\beta}(t) = \frac{1}{T} \sum\limits_{\tau} \langle \mathcal{O}_\alpha(t+\tau)\mathcal{O}_\beta(\tau)\rangle~~(\alpha,\beta=1,2,\ldots, 7),
\end{equation}
\end{widetext}
where the correlation functions with different source time slices $\tau$ are averaged to increase the statistics.  Figure~\ref{fig:SM_schematic-diagram} shows the quark diagrams after Wick's contraction that are involved in the calculation of $C_{\alpha\beta}(t)$. All the connected quark diagrams and the diagrams including the light quark annihilation effects are taken into account, while the charm quark annihilation effects are neglected owing to the OZI rule.

First, we take the following procedure to check the relevance of individual operators $\mathcal{O}_\alpha\in \mathcal{S}$ to the $D\bar{D}^*$
scattering we are interested in:
\begin{itemize}
    \item We start with the subset $\{\mathcal{O}_1,\mathcal{O}_2,\mathcal{O}_3\}$ ($c\bar{c}$ operators) of $\mathcal{S}$, whose correlation matrix $\{C_{ij}(t),i,j=1,2,3\}$ is actually a submatrix $\{C_{ij},i,j=1,2,3\}$ of $C_{\alpha\beta}(t)$. We solve the generalized eigenvalue problem (GEVP) to the submatrices of $C_{ij}(t)$,
          \begin{equation}\label{eq:SM-gevp}
              C_{ij}(t_1)v_j^{(n)}(t_1,t_0)=\lambda^{(n)}(t_1,t_0)C_{ij}(t_0)v_j^{(n)}(t_1,t_0),
          \end{equation}
          for given $t_1,t_0$. Thus we obtain the optimized correlation functions $C^{(n)}(t)=v_i^{(n)}(t_1,t_0)v_j^{(n)}(t_1,t_0) C_{ij}(t)$. This procedure runs over the cases at the four $m_\pi$'s. The effective masses of $C^{(1)}(t)$ and $C^{(2)}(t)$ are illustrated in the rightmost column (column (e)) in each panel of Fig.~\ref{fig:SM_effmass_gevp}. When
          $\mathcal{O}_7(J/\psi\omega,q=0)$ is added to the operator subset, the solution to the corresponding GEVP gives an additional energy level close to the $J/\psi \omega$ threshold, as shown in column (c) as black points in each panel of Fig.~\ref{fig:SM_effmass_gevp}.
    \item Now we perform the GEVP analysis to the operator subset $\{\mathcal{O}_4,\mathcal{O}_5,\mathcal{O}_6, \mathcal{O}_7\}$. The effective masses of the four optimized correlation functions are plotted in column (d) of each panel of Fig.~\ref{fig:SM_effmass_gevp}, where the effective mass illustrated by black points is almost the same as that in column (c). Other three effective masses have good signal-to-noise ratios but do not show plateaus even in the time range $t/a_t\in[25,40]$.
    \item If the GEVP analysis is performed to the full operator set $\mathcal{S}$, we obtain seven optimized correlation functions. The effective masses of the lowest six states
          are plotted in column (a) of each panel of Fig.~\ref{fig:SM_effmass_gevp}. One can see that the effective masses of the lowest five states tend to saturate on plateaus. Column (b) in each panel of Fig.~\ref{fig:SM_effmass_gevp} shows the effective masses when the $J/\psi\omega$ operator is excluded from the operator set. In comparison with column (a),
          the state illustrated by black points disappears but other states are still there and almost do not change.

\end{itemize}

The observations above indicate that the $J/\psi\omega$ operator ($\mathcal{O}_7$) almost does not couple with $c\bar{c}$ and $D\bar{D}^*$ operators and are therefore nearly irrelevant to the $D\bar{D}^*$ scattering. In contrast, $c\bar{c}$ and $D\bar{D}^*$ operators strongly couple with each other and must be considered together. These observations are in agreement with the previous work Ref.~\cite{Prelovsek:2013cra}.

As for the energy levels $E_n$ to be determined, we take the convention that the energy levels reflected by the effective masses in column (b) of Fig.~\ref{fig:SM_effmass_gevp} are ordered as $E_1, E_2,\ldots, E_6$ from bottom to top. For a specific subset of the full operator set $\mathcal{S}$, the energy level that is adjacent to $E_n$ has the same index $n$. We do not include the energy level (indicated by black points) corresponding to the $J/\psi\omega$ to the level order.

For each specific operator subset, we first perform two-exponential fits to the optimized correlation functions $C^{(n)}(t)$ introduced above and take the lowest energy to be an estimate of $E_n$, which are listed in Tables~\ref{tab:energies-M245}-\ref{tab:energies-M415} for all the four $m_\pi$'s. In the data blocks in each table, the first two rows of the leftmost block are almost the same and are the energy levels $E_n$ from $c\bar{c}$ operators with and without the $J/\psi\omega$ operator. The energies $E_n$ in the first two rows of the middle block are from $D\bar{D}^*$ operators with and without the $J/\psi\omega$ operator. The energies $E_n$ in the last two rows are from both $c\bar{c}$ and $D\bar{D}^*$ operators (with and without the $J/\psi\omega$ operator. The energy values in bold are determined from the ratio method (see below). It is seen that the inclusion or exclusion of the $J/\psi\omega$ operator almost does not affect other energy values.  So we exclude the $J/\psi\omega$ operator in the following discussions.

Comparing the energy levels from only $c\bar{c}$ operators, those from only $D\bar{D}*$ operators, and those from both $c\bar{c}$ and $D\bar{D}^*$ operators, the strong correlation between $c\bar{c}$ and $D\bar{D}^*$ operators is obviously manifested. Especially, the three energy levels from only $D\bar{D}*$ operators are far from non-interacting $D\bar{D}^*$ energies. This is attributed to the light quark annihilation effects illustrated in Fig.~\ref{fig:SM_schematic-diagram}. Therefore, it is necessary to include $c\bar{c}$ operators when
the $I=0$ $D\bar{D}^*$ scattering is considered. In this sense, the energies $E_n$ (the bottom row in Tables~\ref{tab:energies-M245}-\ref{tab:energies-M415}) from both $c\bar{c}$ and $D\bar{D}^*$ operators
are taken as good estimates of the eigenvalues of the lattice Hamiltonian relevant to the $I^GJ^{PC}=0^+1^{++}$ charmonium-like system.

\begin{figure*}[t]
    \centering
        \includegraphics[width=1\linewidth]{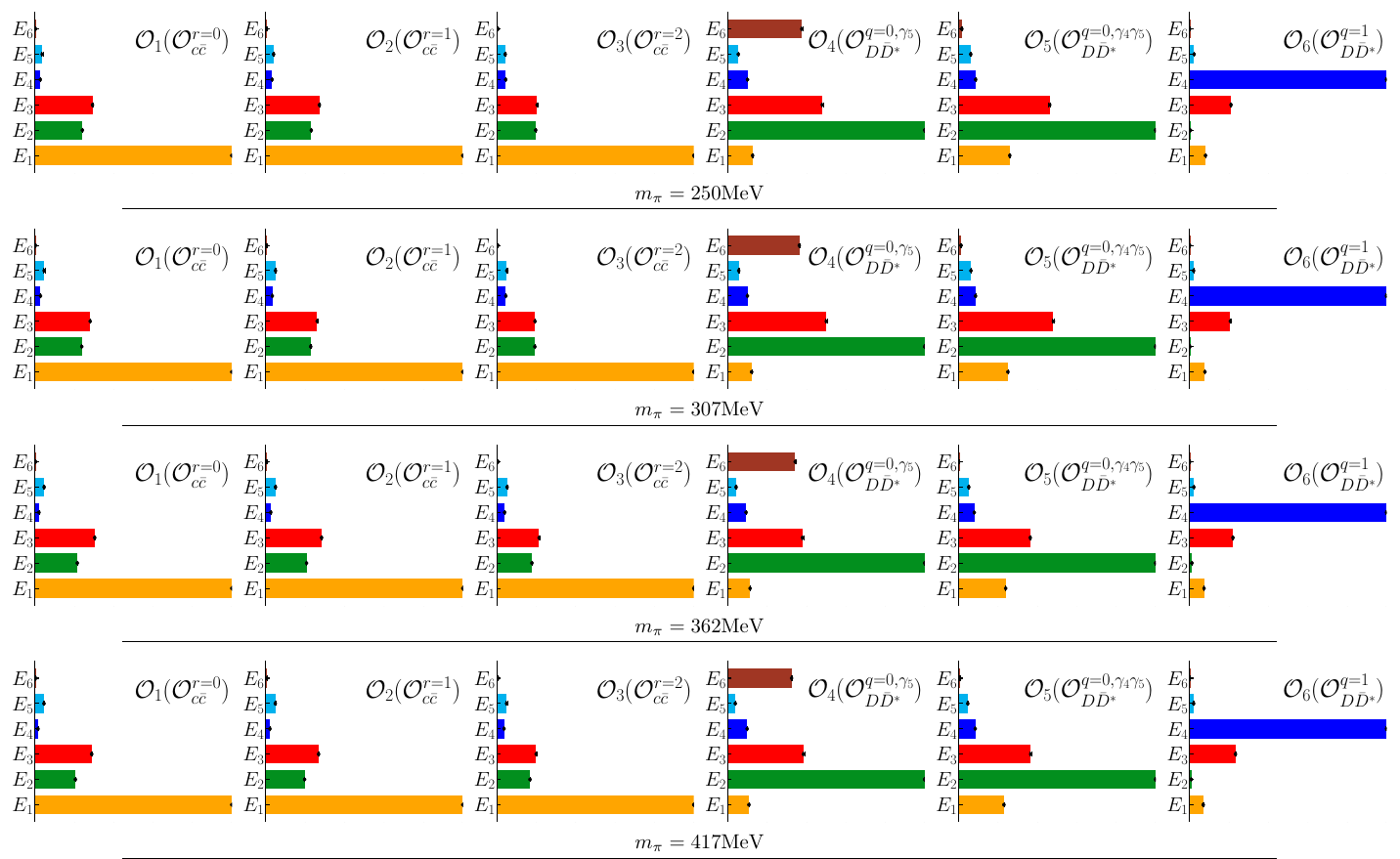}
    \caption{Relative couplings $Z_\alpha^{(n)}=\langle 0|\mathcal{O}_\alpha|n\rangle$ at various $m_\pi$. For each operator $\mathcal{O}_\alpha$, $Z_\alpha^{(n)}$ is normalized by the largest value in $\{Z_\alpha^{(n)}, n=1,2,\ldots, 6\}$. For each state $|n\rangle$, $\{Z_\alpha^{(n)},\alpha=1,2,\ldots, 6\}$ (in the same color) signal the relative importance of $c\bar{c}$ and $D\bar{D}^*$ components.}
    \label{fig:SM-overlap-weight-2}
\end{figure*}

To be specific, we determine the finite volume energies based on the correlation matrix of $c\bar{c}$ and $D\bar{D}^*$ operators, which compose an operator set $\tilde{\mathcal{S}}=\{\mathcal{O}_{\alpha}| \alpha = 1,\cdots, 6\} = \{ \mathcal{O}_{c\bar{c}}^{r=0}, \mathcal{O}_{c\bar{c}}^{r=1}, \mathcal{O}_{c\bar{c}}^{r=2}, \mathcal{O}_{D\bar{D}}^{\mathbf{q}=0, \gamma_5}, \mathcal{O}_{D\bar{D}}^{{q}=0, \gamma_4\gamma_5}, \mathcal{O}_{D\bar{D}}^{{q}=1}\}$. By solving the GEVP to the corresponding correlation matrix $C_{\alpha\beta}(t)$ similarly to Eq.~(\ref{eq:SM-gevp}), we obtain six eigenvectors $\{v_\alpha^{(n)},n=1,2,\ldots, 6\}$ and six optimized correlation functions $\{C^{(n)}(t)=v_\alpha^{(n)}v_\beta^{(n)}C_{\alpha\beta}(t), n=1,2,\ldots,6\}$. It is expected that $C^{(n)}(t)$ is contributed most from the $n$-th eigenstate with the eigen-energy $E_n$ of the lattice Hamiltonian. The physical status of these states can be explored qualitatively by the coupling factor $Z_\alpha^{(n)}$ of $\mathcal{O}_\alpha$ to the $n$-th state, which is extracted through the expression
\begin{equation}\label{eq:overlap-factors}
    \begin{aligned}
        Z_\alpha^{(n)} \equiv |\langle 0 |\mathcal{O}_\alpha|n\rangle|
        \approx & \frac{
            \sqrt{2 E_n} C_{\alpha\beta}(t^*) v_\beta^{(n)}}
        {\sqrt{
                v_\beta ^{(n)}v_\gamma^{(n)} C_{\beta\gamma}(t^*)  e^{-E_n t^*}
            }}
    \end{aligned}
\end{equation}
at $t^*$ where the effective mass plateaus are almost reached. The relative coupling factors $Z_\alpha^{(n)}$ (normalized by the maximal value of $Z_\alpha^{(n)}$ for each operator $\mathcal{O}_\alpha$) are shown in Fig.~\ref{fig:SM-overlap-weight-2}.
Obviously, the $c\bar{c}$ operators $\mathcal{O}_{1,2,3}$ couple most to the $E_1$ state and also have substantial overlaps to the $E_2$ and $E_3$ states. In contrast, $\mathcal{O}_{4,5}$ ($D\bar{D}^*$ operators with $q=0$) couple mainly to $E_2$, and also overlap substantially to $E_1$ and $E_3$ states. $\mathcal{O}_{4}$ also has a large overlap to $E_6$.
$\mathcal{O}_6$ ($\mathcal{O}_{D\bar{D}^*}(q=1)$ operator) couples predominantly to $E_4$ and has a sizable coupling to the $E_3$ state. Based on these information, we make the following raw
identification:
\begin{itemize}
    \item The $E_1$ state of an energy around $3.5$ GeV
          is unambiguously assigned to be the conventional $\chi_{c1}(1P)$ state.
    \item The $E_6$ state is coupled mainly by the $D\bar{D}^*(q=0)$ operator, but has very high energy around 4.6-4.7 GeV (see the bottom rows of Tables~\ref{tab:energies-M245}-\ref{tab:energies-M415}). According to the experimental evidence the mass splitting of $1S-2S$
          mesons are usually around 600 MeV, the $E_6$ state is very likely composed of a ground $D^{(*)}$ state and an excited $\bar{D}^{(*)}$ state with relative zero momentum.
    \item The values of $E_5$ have fairly large errors and are insensitive to the inclusion or exclusion of $D\bar{D}^*$ operator. On the other hand, $E_5$ is relatively far from the energy region relevant to $X(3872)$, so we do not pay much attention to it.
    \item The $E_2,E_3,E_4$ states are very relevant to $D\bar{D}^*$ scattering and $X(3872)$, therefore it is very crucial for them to be
          determined as precisely as possible.
\end{itemize}
\begin{figure*}[t]
    \centering
    \includegraphics[width=0.49\linewidth]{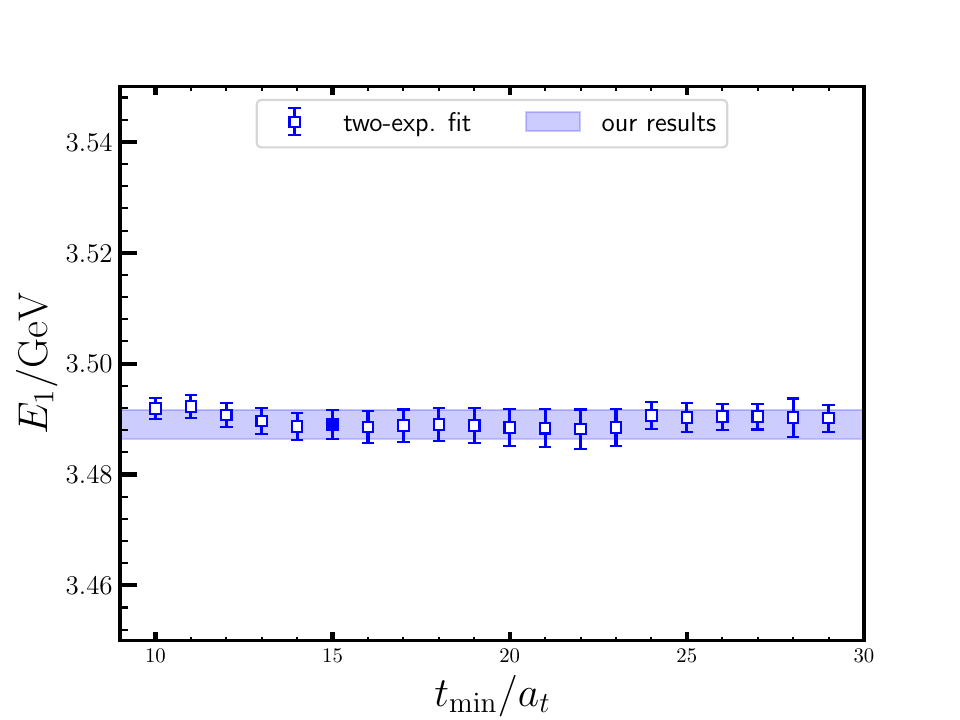}
    \includegraphics[width=0.49\linewidth]{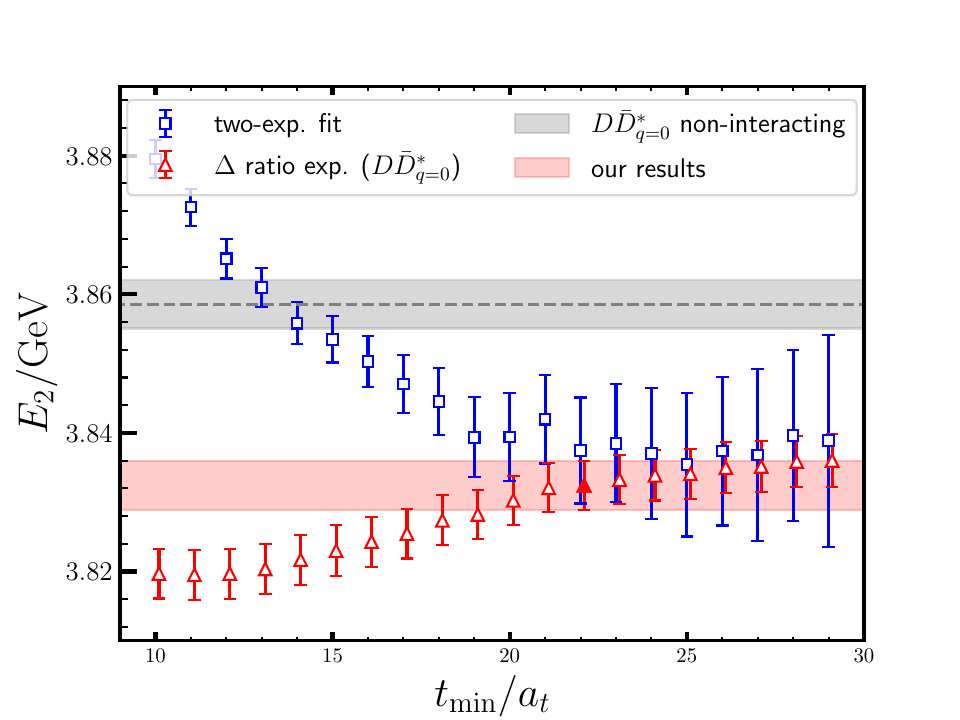}
    \includegraphics[width=0.49\linewidth]{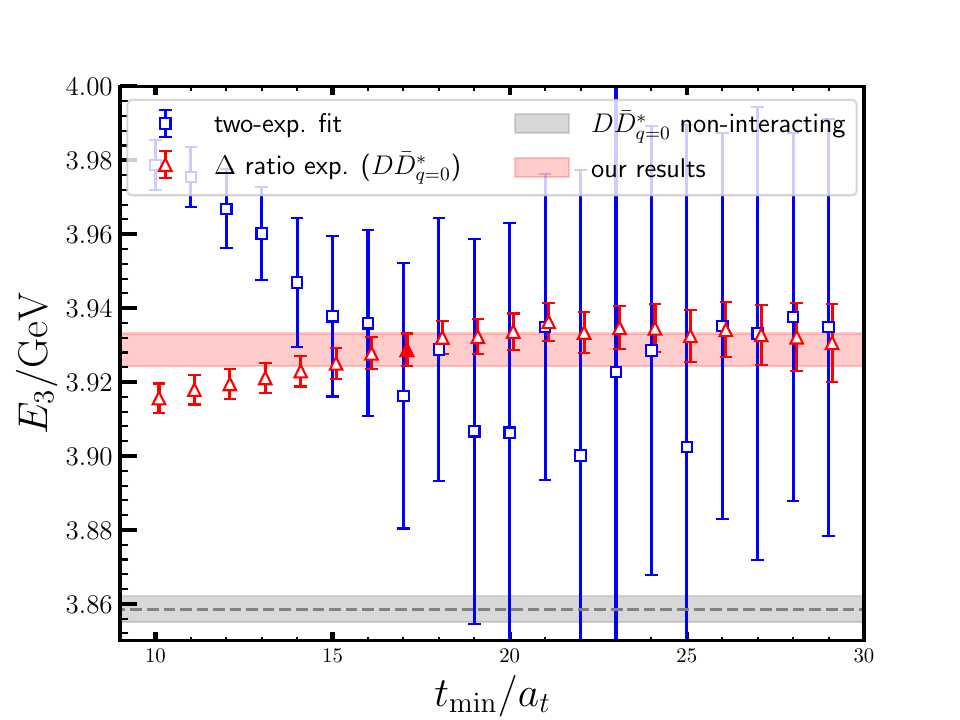}
    \includegraphics[width=0.49\linewidth]{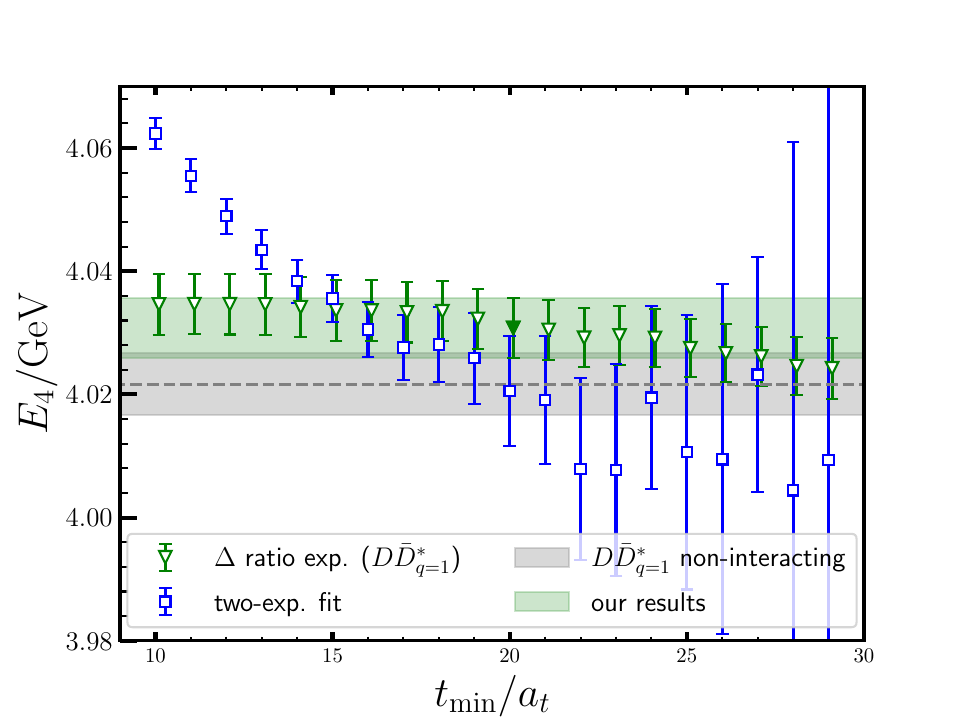}
    \caption{The self-consistency check of the fitted $E_{1,2,3,4}$ through different methods (Fit A, Fit B or Fit C in Eq.~\ref{eq:SM-fitscheme}) at $m_\pi=250$ MeV. The $x$-axis $t_\mathrm{min}$ denotes the different lower bounds of the fit window $[t_\mathrm{min},t_\mathrm{max}]$ ({the upper bound is fixed to be $t_\mathrm{max}=35$}).
        {\bf Top left}: the fitting results of $E_1$ at different $t_\mathrm{min}$. The filled point indicates our final result. 
        {\bf Top right}: $E_2$ results. The grey band illustrates the non-interacting $D\bar{D}^*$ energy $E_{D\bar{D}^*}^{q=0}$ (the $D\bar{D}^*$ threshold). The blue points show the two-exponential fit (Fit A) results of $E_2$, while the red points are the results $E_2=E_{D\bar{D}^*}^{q=0}+\Delta_2(0)$ from the ratio method (Fit B). 
    The results of fit A and Fit B are consistent with each other for $t_\mathrm{min}/a_t\in [15,28]$ but the latter have quite small statistical errors. So we take Fit B result at $t_\mathrm{min}/a_t=22$ (indicated by the filled point) as our final result for $E_2$ (the red band). 
    {\bf Bottom left}: $E_3$ results. The figure legends are similar to the case of $E_2$. Although the consistency of the results of Fit A and Fit B, it is seen that the Fit B results have much smaller statistical errors and show little dependence on $t_\mathrm{min}$. So again we take Fit B result at $t_\mathrm{min}/a_t=17$ as our final result for $E_2$ (the red band).
        {\bf Bottom right}: $E_4$ results: The grey band illustrates the non-interacting $D\bar{D}^*$ energy $E_{D\bar{D}^*}^{q=1}$. The blue points illustrate the Fit A result, while the green points are the results from Fit C (the ratio method) with $E_4=E_{D\bar{D}^*}^{q=0}+\Delta_4(1)$. The results of Fit C are consistent with the two-exponential fit (Fit A) results but have much smaller statistical errors. So we take the Fit C results at $t_\mathrm{min}/a_t=20$ as our final results (the green band).
    }
    \label{fig:fit-stability-m245}
\end{figure*}

\begin{figure*}[t]
    \centering
    \includegraphics[width=0.49\linewidth]{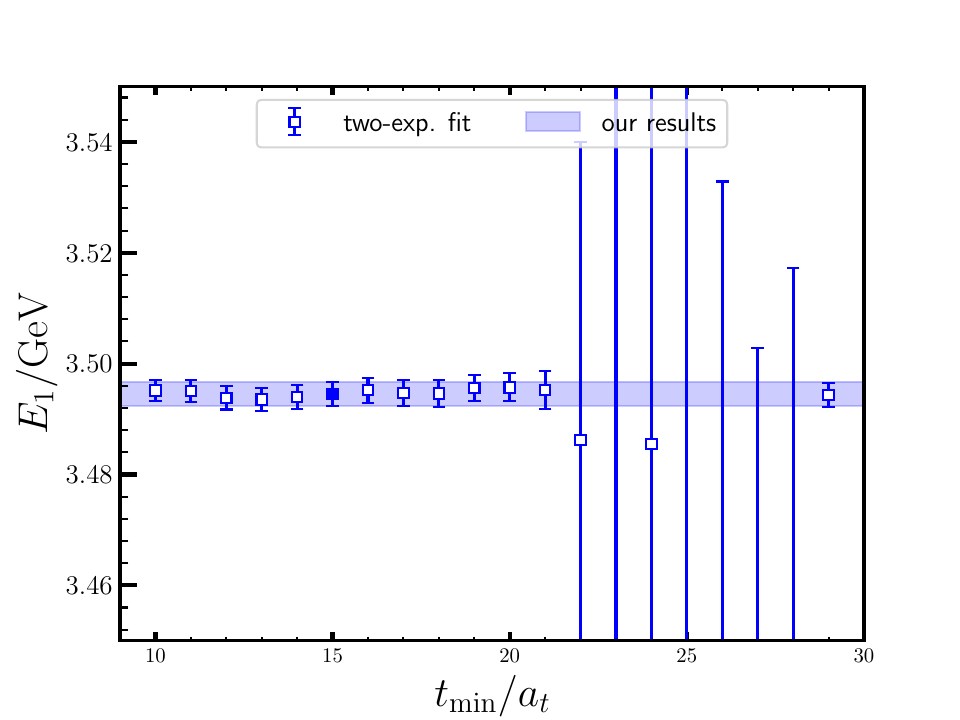}
    \includegraphics[width=0.49\linewidth]{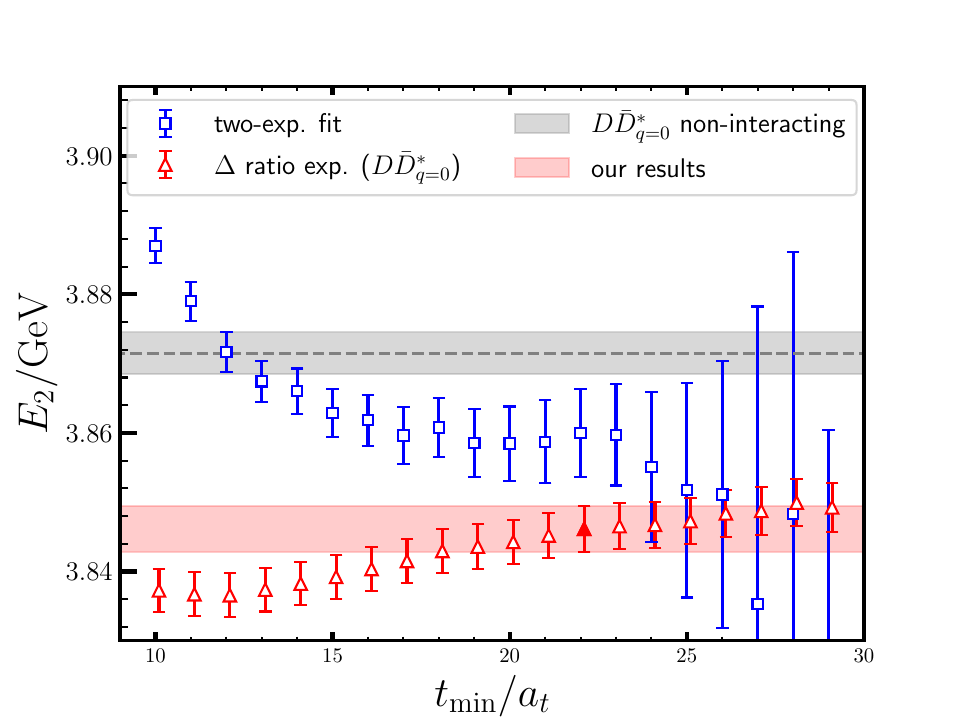}
    \includegraphics[width=0.49\linewidth]{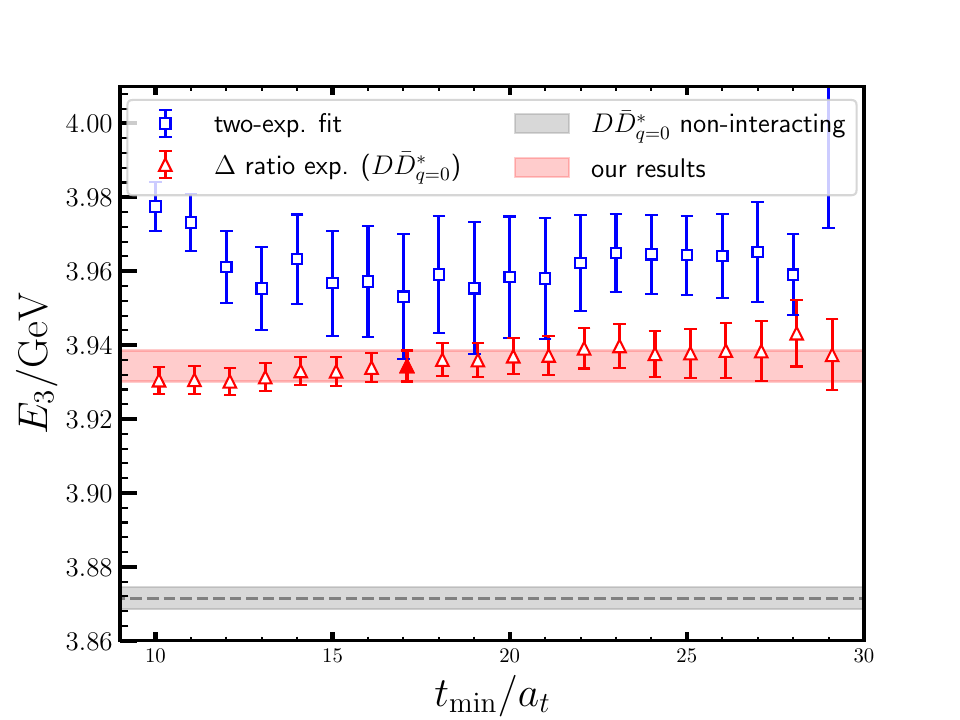}
    \includegraphics[width=0.49\linewidth]{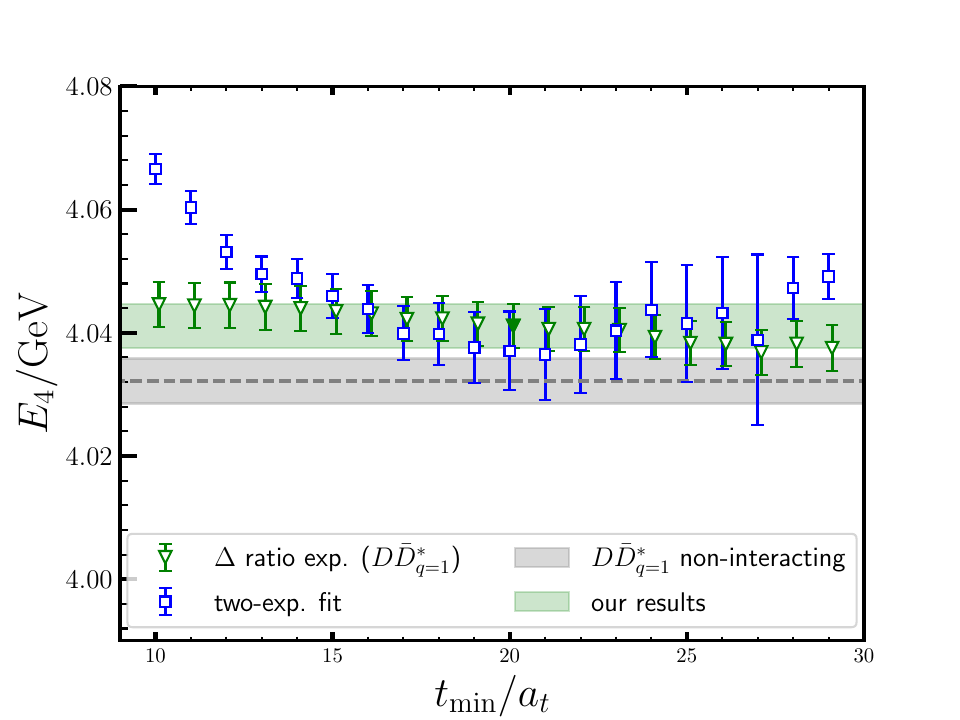}
    \caption{The self-consistency check of the fitted $E_{1,2,3,4}$ through different methods at $m_\pi = 307$ MeV. The layout is similar to Fig.~\ref{fig:fit-stability-m245}. The filled point in each panel indicates the $t_\mathrm{min}/a_t$ where our final result is taken.}
    \label{fig:fit-stability-m305}
\end{figure*}

\begin{figure*}[t]
    \centering
    \includegraphics[width=0.49\linewidth]{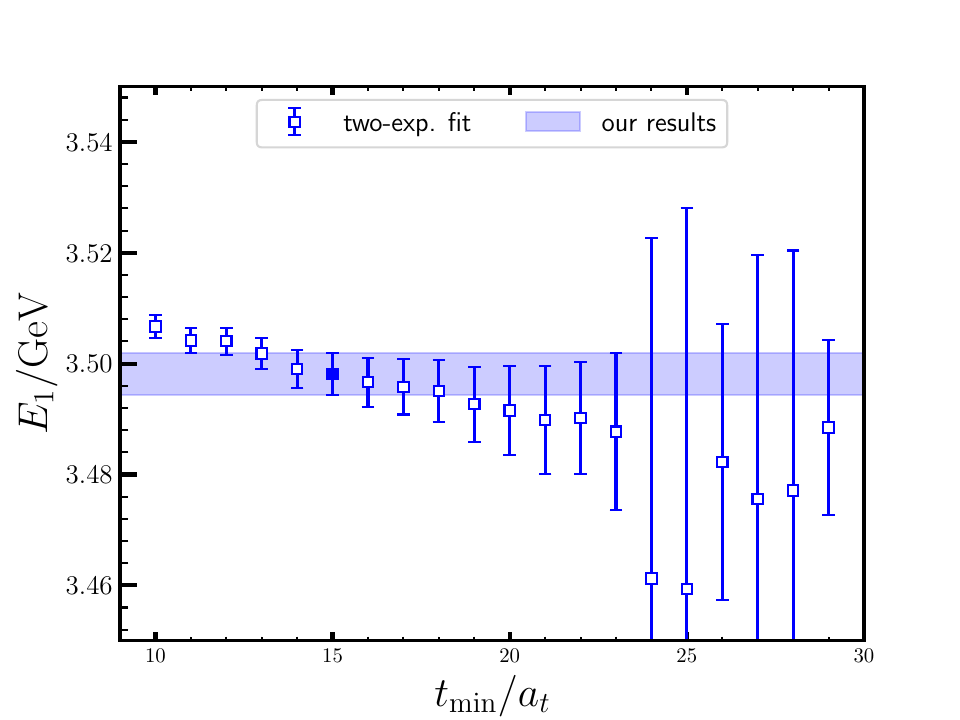}
    \includegraphics[width=0.49\linewidth]{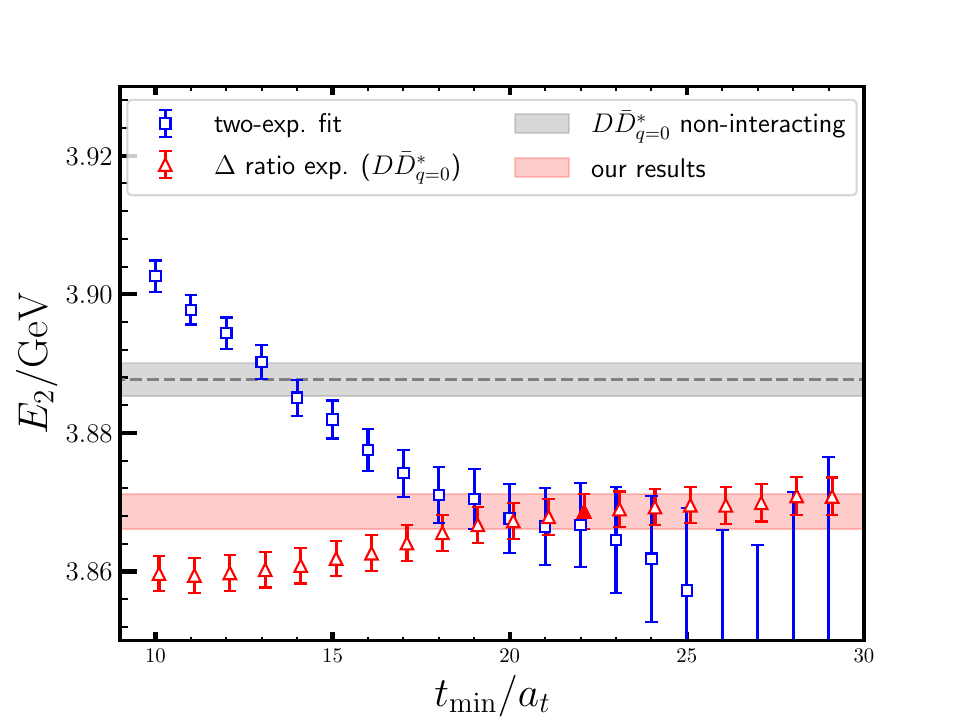}
    \includegraphics[width=0.49\linewidth]{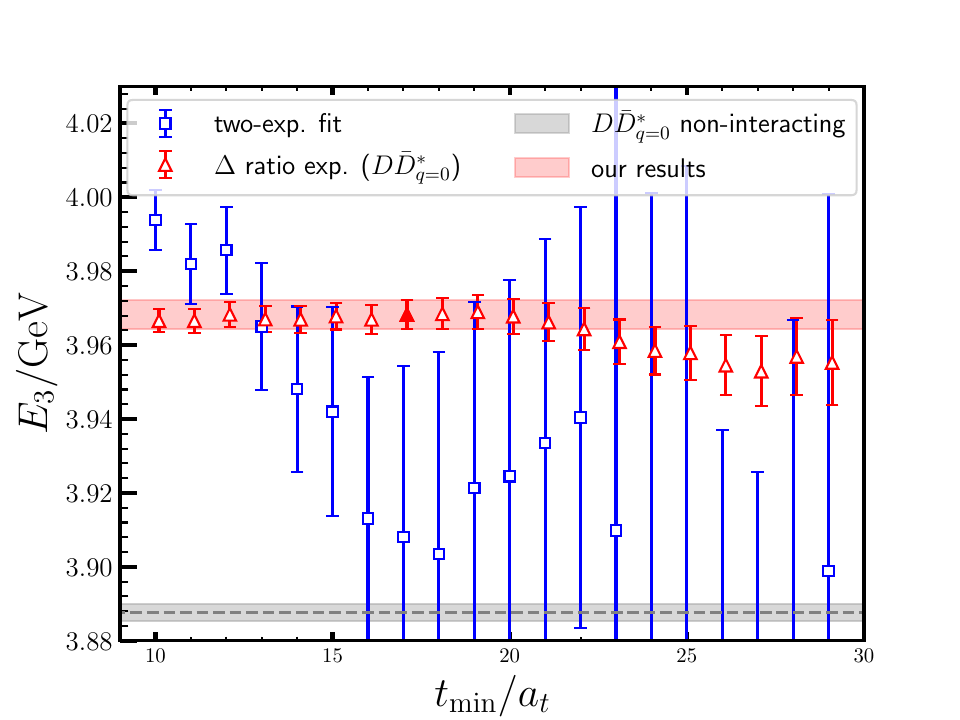}
    \includegraphics[width=0.49\linewidth]{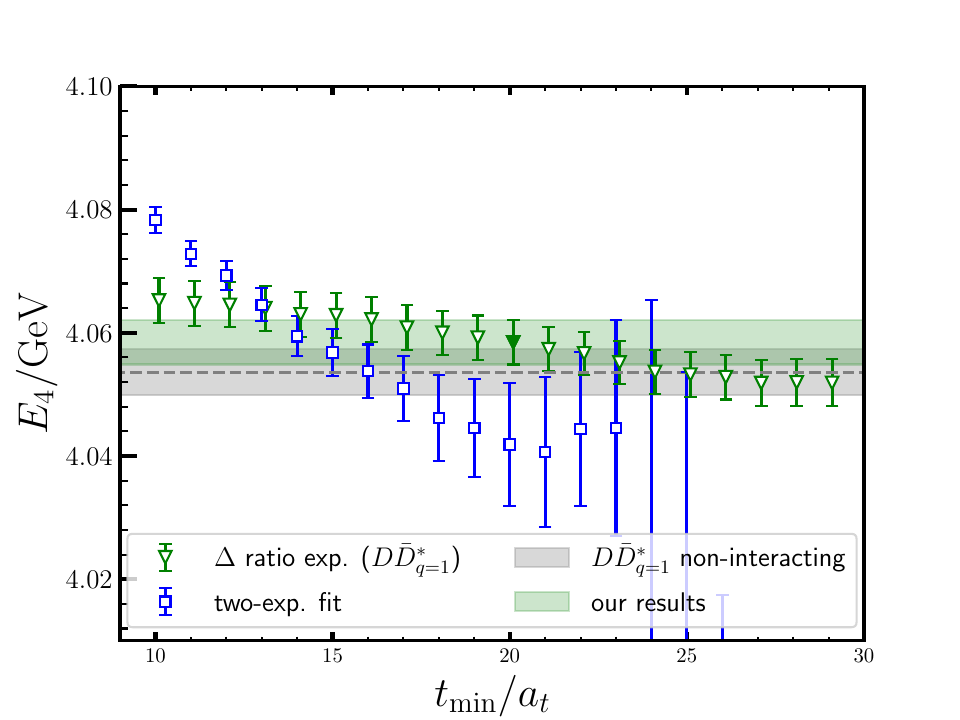}
    \caption{The self-consistency check of the fitted $E_{1,2,3,4}$ through different methods at $m_\pi = 362$ MeV. The layout is similar to Fig.~\ref{fig:fit-stability-m245}. The filled point in each panel indicates the $t_\mathrm{min}/a_t$ where our final result is taken.
    }
    \label{fig:fit-stability-m360}
\end{figure*}

\begin{figure*}[t]
    \centering
    \includegraphics[width=0.49\linewidth]{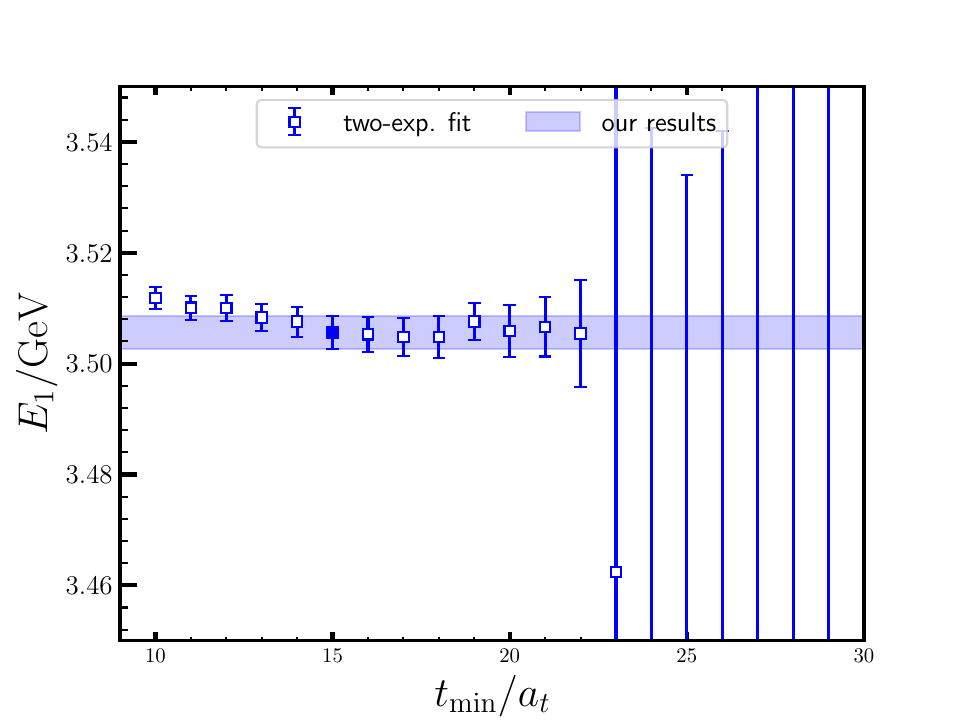}
    \includegraphics[width=0.49\linewidth]{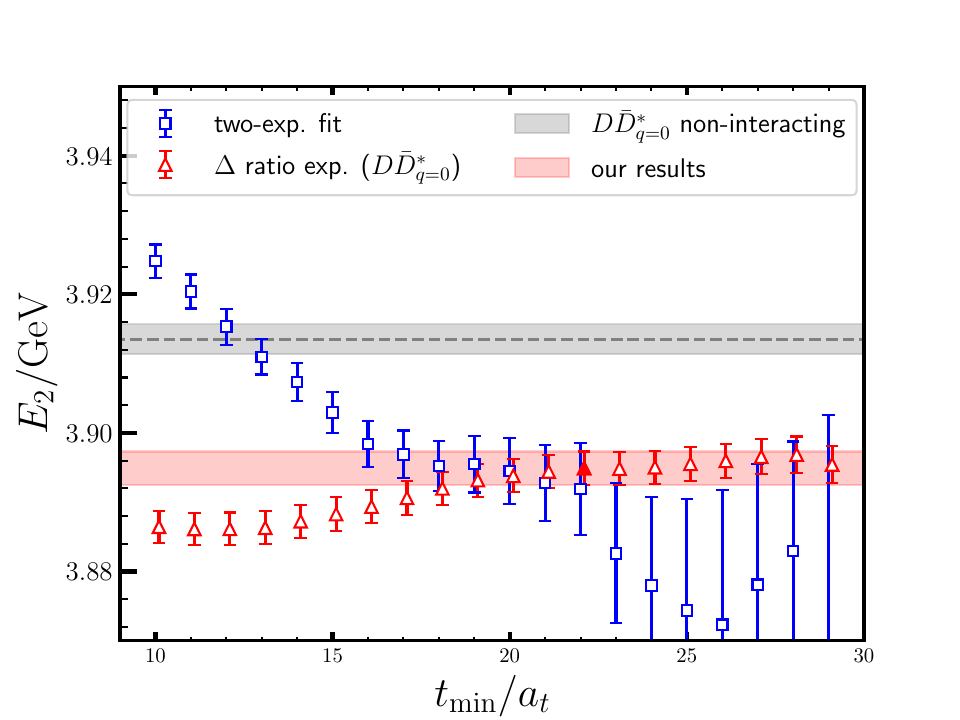}
    \includegraphics[width=0.49\linewidth]{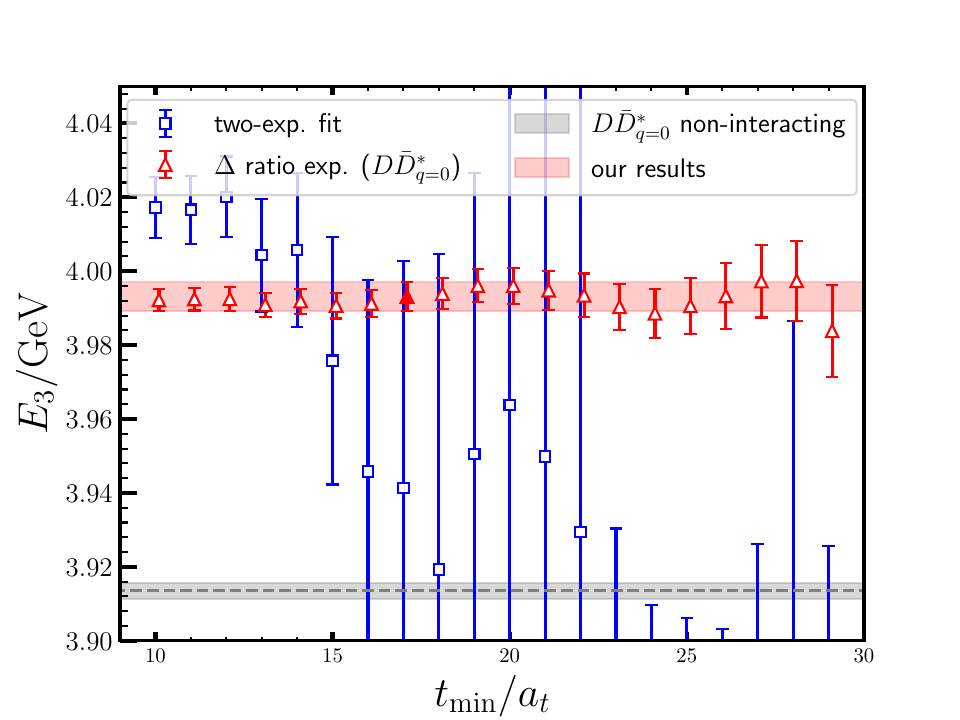}
    \includegraphics[width=0.49\linewidth]{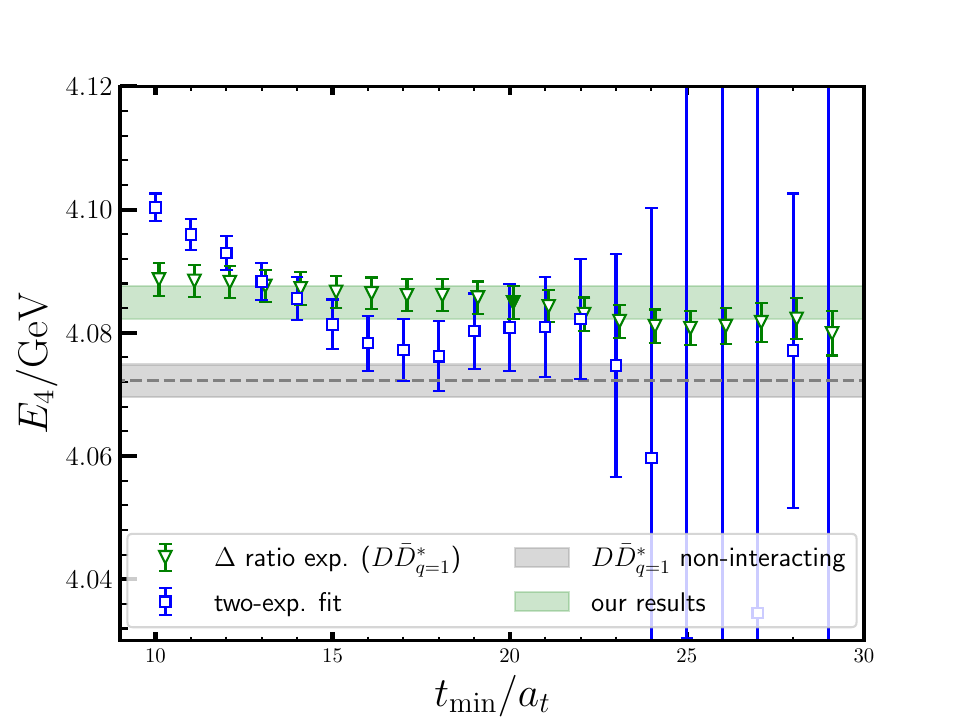}
    \caption{The self-consistency check of the fitted $E_{1,2,3,4}$ through different methods at $m_\pi = 417$ MeV. The layout is similar to Fig.~\ref{fig:fit-stability-m245}.  The filled point in each panel indicates the $t_\mathrm{min}/a_t$ where our final result is taken.}
    \label{fig:fit-stability-m415}
\end{figure*}

\setcounter{figure}{0}
\renewcommand{\thefigure}{C\arabic{figure}}
\setcounter{table}{0}
\renewcommand{\thetable}{C\arabic{table}}
\section{Systematic uncertainties from anisotropic lattice}\label{sec:appendix-S3}
To facilitate a clearer comprehension of physical values in this study, the fixed $a_t^{-1}$ mean values are adopted to present the finite volume energies, notwithstanding the uncertainties associated with the lattice space $a_s$ and the anisotropy parameter $\xi\equiv a_s/ a_t$.
Besides, a systematic uncertainty on anisotropic parameter $\xi$ should be taken into account in finite volume scattering analysis.
Generally, the finite volume scattering momenta are determined by solving the dispersion relation from prescribed energy levels,
\begin{equation}\label{eq:SM_scattering_momemta}
    \begin{aligned}
        E_n =         & \sqrt{m_D^2 + \vec{p}_n^2}+\sqrt{m_{D^*}^2 + \vec{p}_n^2},                                         \\
        \vec{p}_n^2 = & \frac{1}{4} E_n^2 -  \frac{1}{2} (m_{D}^2 + m_{D^*}^2) + \frac{(m_{D}^2 - m_{D^*}^2)^2}{ 4 E_n^2},
    \end{aligned}
\end{equation}
So we use several fitting methods to extract the values of $E_{2,3,4}$ and
compare the results to check the self-consistency:
\begin{eqnarray}\label{eq:SM-fitscheme}
    \mathrm{Fit~A}:&&C^{(n)}(t)=(A_1 e^{-E_n t}+A_2 e^{-E't})\nonumber\\
    &&~~~~~~~~~~~~~~~~~+(t\to (T-t)),\nonumber\\
    \mathrm{Fit~B}:&&R^{(n)}(t,q=0)=\frac{C^{(n)}(t)}{C_D(t,q=0)C_{D*}(t,q=0)}\nonumber\\
    &&~~~~~~~~~~~~~~~~~\approx A e^{-\Delta_n(0) t},\nonumber\\
    \mathrm{Fit~C}: &&R^{(n)}(t.q=1)=\frac{C^{(n)}(t)}{C_D(t,q=1)C_{D*}(t,q=1}\nonumber\\
    &&~~~~~~~~~~~~~~~~~\approx A e^{-\Delta_n(1) t}.
\end{eqnarray}
Fit A is a two-state fitting. Fit B involves fitting the two-point corrector ratio $R(t)$ that $C(t)$ over the non-interacting $C_D(t,q=0)$ and $C_{D*}(t,q=0)$ using a single exponential function and then adding its non-interacting energy levels $E_n=E_{D\bar{D}^*}^{q}+\Delta_n(q)$ through the jackknife analysis. Fit C is similar to Fit B but accounts for the non-interacting $C_D(t,q=1)$ and $C_{D*}(t,q=1)$.
The comparisons are illustrated in Figs.~\ref{fig:fit-stability-m245}-\ref{fig:fit-stability-m415} for the four $m_\pi$'s. In each figure, the top-left panel shows the fitted results of $E_1$ at different $t_\mathrm{min}$ of the fit window $[t_\mathrm{min},t_\mathrm{max}]$.
The maximum timeslice range $t_\mathrm{max}$ is set to 40 across all fits, as this is in the saturation range of these observed energy levels.

\begin{itemize}
    \item In the case of energy level $E_1$, it is far below the $D\bar{D}^*$ threshold, and the simple two-states fitting performs well enough, while the ratio-exponential fits are unflavored due to the substantial effects of excited states artifacts.
    \item The energy level $E_2$ is slightly below $D\bar{D}^*$ threshold about 20 MeV. The results from Fit A and Fit B reach commendable consistency.
    \item For energy level $E_3$, the results from Fit B show greater precision compared to Fit A (two-state fit). Consequently, the results from Fit B are adopted in our determination.
    \item In the case of energy level $E_4$, Fit B (at relative momentum $q=0$) gives unstable results, as observed the energy level plateau exhibits a gradual slope, which suggests the potential contamination of excited states. Conversely, Fit C is observed to give a more robust and dependable result for the energy plateau.
\end{itemize}

Practically, the dimensionless scattering momenta are derived from the relation $ q^2 = (\frac{2\pi}{\hat{L}\xi})^2 (a_t \vec{p})^2$, where $(a_t \vec{p})^2$ expressed in lattice unit.
In such a relation, it is necessary to take into account the uncertainty of the anisotropy parameter caused by solving the dimensionless scattering momenta.
Since this study focuses on the $D\bar{D}^*$ channel, We choose to employ the anisotropy parameter measured from $D$ and $D^*$ meson.
Given the slight deviation observed in the dispersion relations of the $D$ and $D^*$ mesons, as documented in Table~\ref{tab:SM-anisotropy-parameter}, we incorporate this discrepancy as a source of systematic uncertainty. The determined values $\xi_{\mathrm{sys.}}$ ensure that the one-sigma uncertainty range encompasses these systematic variations.

\begin{table*}[t]
    \caption{Anisotropy parameter $\xi$ and dispersion relation of various ensembles. The derivation of $\xi_D$ and $\xi_{D^*}$ is considered to be a systematic uncertainty in Eq.~(\ref{eq:SM_scattering_momemta}), noted as $\xi_{\mathrm{sys.}}$.}
    \label{tab:SM-anisotropy-parameter}
    \begin{ruledtabular}
        \begin{tabular}{lccccc|c}
            ens. & $m_\pi / \mathrm{MeV}$ & $a_t^{-1}/\mathrm{GeV}$ & $\xi_{J/\psi}$ & $\xi_D$    & $\xi_{D^*}$ & $\xi_{\mathrm{sys.}}$ \\\hline
            M245 & 250(3)                 & $7.276$                 & 5.061 (10)     & 5.070 (22) & 5.126 (54)  & 5.11(7)               \\
            M305 & 307(2)                 & $7.187$                 & 5.083 (10)     & 5.025 (16) & 5.068 (30)  & 5.05(5)               \\
            M360 & 362(1)                 & $7.187$                 & 4.989 (10)     & 4.974 (14) & 4.990 (31)  & 4.99(3)               \\
            M415 & 417(1)                 & $7.219$                 & 5.021 (11)     & 5.031 (16) & 5.091 (39)  & 5.07(6)               \\
        \end{tabular}
    \end{ruledtabular}
\end{table*}